\documentclass[aps,floatfix,showpacs,superscriptaddress, nofootinbib,twocolumn]{revtex4-1} 
\usepackage{longtable,dcolumn,epsfig,color,amsmath, hyperref, placeins}

\begin{document}

\title{Axion emission from nuclear magnetic dipole transitions}
\author{R. Massarczyk}
\email[]{massarczyk@lanl.gov}
\author{P.-H. Chu}
\author{S.R. Elliott}
\affiliation{Los Alamos National Laboratory, Los Alamos, NM, USA}

\date{\today}

\begin{abstract}
Nuclear transitions are one possible source of axions but past searches were restricted to specific transitions. In this manuscript, we propose to extend the search for axions and axion-like particles to more a complex environment that would result in a number of correlated observables. By including creation mechanisms that have their origin in the Carbon-Nitrogen-Oxygen (CNO) cycle, we show that the search for solar axions should not only be restricted to the\,keV-mass region. We discuss limitations, such as the lifetime and the mass, that create a challenge for an Earth-bound experiments. We show that it is possible to use the same creation mechanisms as used in solar axions to search with a comparable rate at nuclear power reactors. 
\end{abstract}

\maketitle

\section{\label{sec:intro}Introduction}

\begin{figure}[t]
\centering
\includegraphics[width=1.0\columnwidth,keepaspectratio=true]{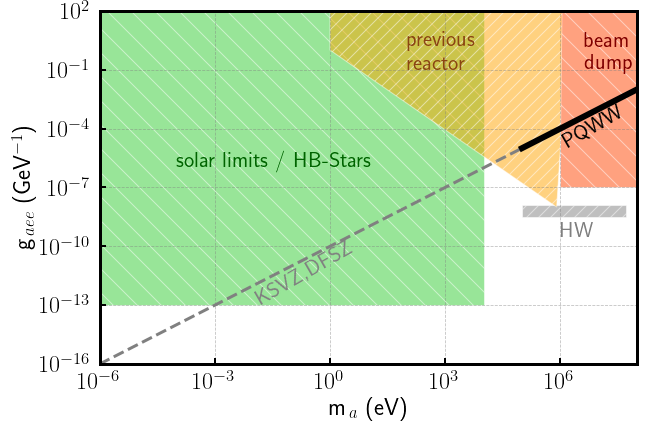}
\includegraphics[width=1.0\columnwidth,keepaspectratio=true]{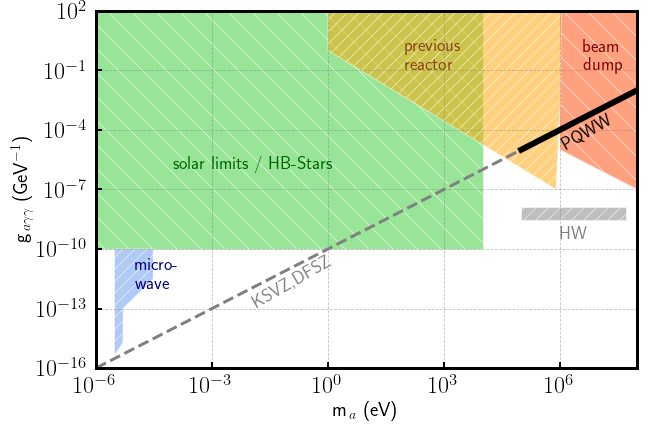}
\caption{(Color figure) Current limits on axion-photon and axion-electron coupling~\cite{PhysRevD.75.052004}. The expectation values for the KSVZ model~\cite{PhysRevLett.43.103}, the DFSZ model~\cite{DINE1981199}, the PQWW model~\cite{PhysRevLett.38.1440}, and HW model~\cite{PhysRevD.70.115001} are shown in black and gray. }
\label{fig:limit}
\end{figure}

Axions are hypothetical particles that were introduced as a solution to the strong CP problem~\cite{PhysRevLett.38.1440, PhysRevLett.40.223, PhysRevLett.40.279}. Numerous models, like the KSVZ model~\cite{PhysRevLett.43.103, SHIFMAN1980493}, the DFSZ model~\cite{DINE1981199, Zhitnitsky:1980tq}, the PQWW model~\cite{PhysRevLett.38.1440}, and the HW model~\cite{PhysRevD.70.115001}, exist. They differ in favored axion mass - coupling parameter space regions, see Fig.~\ref{fig:limit}.
One suggested creation mechanism of axions is through the coupling to nucleons. Axions behave as pseudoscalar particles, which would result in coupling mechanisms similar to photons with magnetic multipole character. Therefore, axions may mediate the relaxation of a nuclear state, that typically decays via a magnetic dipole transition~\cite{PhysRevD.18.1607}. If so, nuclear transitions can be regarded as a potential source of axions. Over the last decade, the search for axions originating from nuclear processes have focused on two possible production sources, solar axions with Earth-bound detectors and axions that produced within strong radioactive sources. 

Solar axion searches often focus on the production by the Primakoff effect~\cite{PhysRevLett.81.5068}. Recent data published by the XENON1T collaboration~\cite{2006.09721} is interpreted by various groups as evidence that the Sun emits axions~\cite{2006.14598,2006.15118, 2006.16220, 2007.05517}. Other authors declare this excess is due to dark matter~\cite{2006.11264, 2006.11938}, or at least suggest it to be non-axionic~\cite{2006.12487} since the deduced axion phase space is in conflict with other bounds. 

In addition to the Primakoff effect, solar axions can be generated through a nuclear based production mechanism from a magnetic dipole transitions. In the Sun, low energy nuclear states can be thermally excited, though the number density of candidate isotopes with low-energy magnetic dipole states is limited within the solar environment. After excitation within the thermal bath, excited isotopes decay quickly back to the ground state. In this deexcitation, a branching between axion and photon emission $\Gamma_a / \Gamma_\gamma$ is possible~\cite{PhysRevD.18.1607}. Under the assumption that a low-mass axion is emitted, this mechanism creates a mono-energetic flux of axions with an intensity comparable to the Primakoff production mechanism~\cite{PhysRevLett.75.3222}. Over the last decade, this approach has been used to search for a 14.4-keV solar axion emitted by $^{57}$Fe~\cite{NAMBA2007398, collaboration_2009, Derbin2011, IOP2013, Armengaud_2013, PhysRevLett.118.161801, III_2018} or $^{83}$Kr~\cite{Gavrilyuk2015}. Using the same physics, axions should be observable from other reactions occurring during the solar burning. A focus on these reactions is motivated by the fact that the concentration of hydrogen and helium are orders of magnitude higher than for medium mass isotopes like $^{57}$Fe or $^{83}$Kr. The parameters needed to estimate the flux, including isotope yields, isotope distributions, as well as reaction rates are given by solar neutrino physics and the detailed solar models. Refs.~\cite{RevModPhys.64.885, Bahcall_2005} review solar neutrino production. 

Another strong motivation is that the total energy of such axions is in the MeV range, far away from the low energy part of measured spectra that are usually heavily contaminated by contributions from natural radioactivity. Previous studies discussed the production of solar axions in the\,keV to\,MeV mass region using the $p(d,\gamma)^3$He-reaction~\cite{RAFFELT1982323, collaboration_2010,PhysRevD.85.092003, Derbin2010, Derbin2013} as well within the creation of $^7$Li~\cite{ collaboration_2010}. While axions within this mass range are not typically considered as dark matter candidates, this mass window has not been fully explored, cf. Fig.~\ref{fig:limit}. However, recent publications~\cite{2004.01193, 2004.08399} discussed the importance of this window as a transition point where laboratory searches and cosmological sensitivities meet. 

In this manuscript we extend the approach to additional reactions within the Sun, and in particular we discuss the possibility of axion emission from transitions within the Carbon-Nitrogen-Oxygen (CNO) cycle. We then use the developed methodology to consider transitions in reactor fission products. As shown in Fig.~\ref{fig:limit}, a number experiments have excluded certain axion models above the\,keV scale~\cite{TURNER199067}; however, alternative models like Ref.~\cite{PhysRevD.70.115001} or Ref.~\cite{Alves:2017avw} show it is possible that such axion-like particles (ALP) exist and survive the current constraints. Due to their limited lifetime, the observation of solar axions with mass greater than 1~MeV is impossible. In Sec.~\ref{sec:axiontime} we will discuss this limitation, and will show that this boundary is in fact even lower. While a number of searches and theoretical perspectives~\cite{IRASTORZA201889} focus on cosmogenic origin, the search for axions in magnetic dipole transitions can be extended to terrestrial sources. Using an Earth-bound source, short flight paths help to avoid the lifetime constraint. A number of light-shine-through-wall experiments~\cite{IRASTORZA201889} have been performed in the past, mainly focusing on axion masses ($m_a$) below 1 eV, motivated by dark matter searches. However, a heavy axion or ALP can act as a potential mediator to a beyond the standard model sector and is not yet excluded. Such particles might be emitted during nuclear decays. Similar to the decays in a solar environment, a branching between magnetic dipole photon and axion production is possible when a nucleus relaxes to a lower energetic state, for example, $^{65}$Cu~\cite{PhysRevD.37.618}, $^{125m}$Te~\cite{Derbin1997} and $^{139}$La~\cite{PhysRevLett.71.4120} radioactive source measurements have set first limits. Due to limitations in source activities these constraints were not competitive. However, one intriguing terrestrial environment in which numerous nuclear decays occur with high rates are nuclear reactors. First studies looked for axions from only a few selected reactor isotopes~\cite{Altmann1995,PhysRevD.75.052004, Chang_2008} or only discussed the production of ALPs without nuclear interactions~\cite{PhysRevLett.124.211804}. In this work, we show that similar to the solar environment, the approach should be extended to consider the numerous possible transitions available within a high power reactor. Since the fission of uranium creates a varied collection of isotopes, we consider a spectral analysis instead of focusing on individual peak signatures. 

\section{\label{sec:axionphoton}Axionic deexcitation of nuclear states}

\begin{figure}[t]
\centering
\includegraphics[width=1.0\columnwidth,keepaspectratio=true]{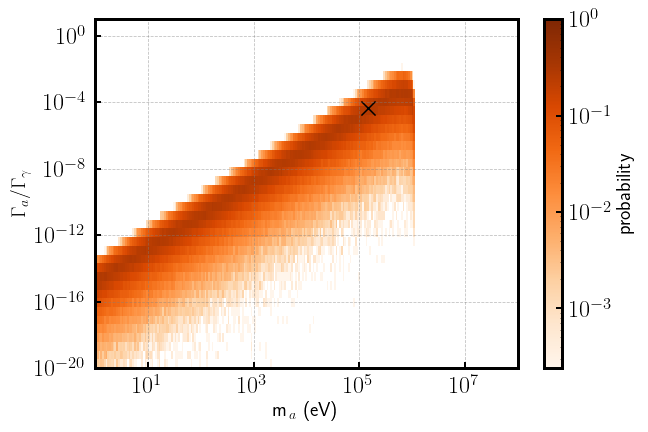}
\caption{(Color figure) Axion-to-photon branching for a 1.115-MeV magnetic dipole transition in $^{65}$Cu~\cite{PhysRevD.37.618} as a function of axion mass calculated using Eq.~\ref{Eq:branching}. For illustration, the nuclear parameters are randomized in each mass bin. The probability functions are normalized for each individual axion-mass bin. The cross marks the branching calculated in Ref. \cite{PhysRevD.37.618}.}
\label{fig:branching}
\end{figure}

The branching ratio for an axionic nuclear deexcitation can be calculated using the photon decay width, $\Gamma_\gamma$, and the width for the axion $\Gamma_a$. With the assumption that $\Gamma_a \ll \Gamma_\gamma$, the probability for axionic emission~\cite{PhysRevD.37.618} can be written as:
\begin{equation}
\label{Eq:1}
 \frac{\Gamma_a}{\Gamma_\gamma} = \frac{1}{2\pi\alpha} \Big(\frac{k_a}{k_\gamma}\Big)^3 \frac{1}{1+\delta^2} \Big(\frac{g^0_{aNN}\beta + g^1_{aNN}}{(\mu_0-1/2)\beta+\mu_1-\eta}\Big)^2
\end{equation}
Here, $\alpha$ denotes the fine-structure constant, 1/137. In the existing literature, the momenta of axion ($k_a$) and photon ($k_\gamma$) are often set to be equal due to the small axion masses considered. In this work, we account for the axion mass by using the relativistic mass-energy relation. The mixing ratio between magnetic dipole ($M1$) and electric quadrupole ($E2$) for a specific transition is represented by the factor $\delta$. $\mu_0=\mu_p+\mu_n=0.88$ and $\mu_1=\mu_p-\mu_n=4.71$ are the isoscalar and isovector magnetic moments, respectively~\cite{PhysRevD.37.618}. The parameters $\beta$ and $\eta$ are two nuclear structure dependent values that will be discussed in the next subsection. The axion nucleon coupling strengths $g^0_{aNN}$ and $g^1_{aNN}$ ~\cite{SREDNICKI1985689,KAPLAN1985215} are defined as:
\begin{equation}
\label{Eq:2a}
 g^0_{aNN} = -7.8\times10^{-8} \frac{6.2\times 10^6}{f_a/\mathrm{GeV}} \Big(\frac{3F - D + 2S}{3}\Big)
\end{equation}
and
\begin{equation}
\label{Eq:2b}
 g^1_{aNN} = -7.8\times10^{-8} \frac{6.2\times 10^6}{f_a/\mathrm{GeV}} \Big((D + F)\frac{1-z}{1+z}\Big).
\end{equation}
We follow the relation for the classical QCD-axion for which this coupling strength and mass are correlated. The axion coupling strength $f_a$ can be defined relative to properties of the pion, as well as the up-down quark mass-ratio $z$, and the mass of the axion itself. 
\begin{equation}
\label{Eq:3}
 f_a = \frac{\sqrt{z}}{1+z}\frac{f_\pi m_\pi}{m_a}
\end{equation}
The parameter $z$ describes the up and down quark mass ratio. The parameters $D$ and $F$ are the invariant matrix elements of the axial current. $S$ is the flavor-singlet axial vector matrix element. For ALP this relation need not hold up and calculations allow $f_a$ and $m_a$ to be independent. Here, we keep these two parameters coupled, so that the axion-mass remains the only free variable, which corresponds to the classic DFSZ model~\cite{DINE1981199}. Similarly, a correlation between both quantities can be also found in the KSVZ model for\,MeV-mass scale axions~\cite{PhysRevLett.38.1440}. Relaxing this relationship opens a new window of phase space. However, as one can see in Fig.~\ref{fig:limit} an analysis with coupled parameters is still valuable, since the experimental limits are still above the coupling-mass limit for certain mass ranges. Using the factors $S$ = 0.68, $z=m_{up}/m_{down}$ = 0.56, $D$ = 0.77, $F$ = 0.48~\cite{DINE1981199,PhysRevLett.75.3222}, $m_\pi$ = 139\,MeV, and $f_\pi$ = 93\,MeV, Eqs.~\ref{Eq:2a}-\ref{Eq:3} can be written as proposed in Ref.~\cite{PhysRevLett.66.2557}:
\begin{align}
 g^0_{aNN} =& -7.8\times10^{-8} \times m_a/\mathrm{eV} \times 0.67,\\
 g^1_{aNN} =& -7.8\times10^{-8} \times m_a/\mathrm{eV} \times 0.35
\end{align}
and
\begin{equation}
 \label{Eq:coupling}
 f_a/\mathrm{GeV} =\frac{6.2 \times 10^6}{m_a/\mathrm{eV}}.
\end{equation}
Restrictions on the parameter $g^1_{aNN}$ are given from Kaon decays~\cite{EICHLER1986101,ARTAMONOV2005192}. In the calculations presented here, we follow the limits proposed in Ref.~\cite{Alves:2017avw}. Under the assumption that the momentum of the transition is completely carried away by an axion with mass $m_a$, the branching can be written as:
\begin{equation}
\label{Eq:branching}
\begin{aligned}
 \frac{\Gamma_a}{\Gamma_\gamma} = & \frac{1}{2\pi\alpha} \Big(\frac{\sqrt{E_\gamma^2-m_a^2}}{E_\gamma}\Big)^3 \frac{1}{1+\delta^2} \\
 & \times \Big(\frac{-7.8\times10^{-8} \times m_a/\mathrm{eV}\, (0.67 \beta + 0.35)}{0.38\beta+4.71-\eta}\Big)^2
\end{aligned}
\end{equation}
Nuclear recoil is a negligible effect for the energies considered here. As shown in Eq.~\ref{Eq:branching}, the branching increases with axion mass, has a maximum just below to the transition energy $E_\gamma$, and vanishes for $m_a \sim E_\gamma$. While the mass of the emitted axions can range from almost zero to energy of the nuclear transition, the best observation range with this approach is in the\,keV to\,MeV mass range. Fig.~\ref{fig:branching} shows the distribution of the branching ratio as a function of axion mass calculated for an 1.115-MeV transition in $^{65}$Cu. For this plot a Monte-Carlo calculation was performed where the parameters $\eta$, $\beta$, and $\delta$ were distributed as discussed in the Sec.~\ref{sec:parameters}. 

\subsection{\label{sec:parameters}Nuclear parameters} 

\begin{figure}[t]
\centering
\includegraphics[width=1.0\columnwidth,keepaspectratio=true]{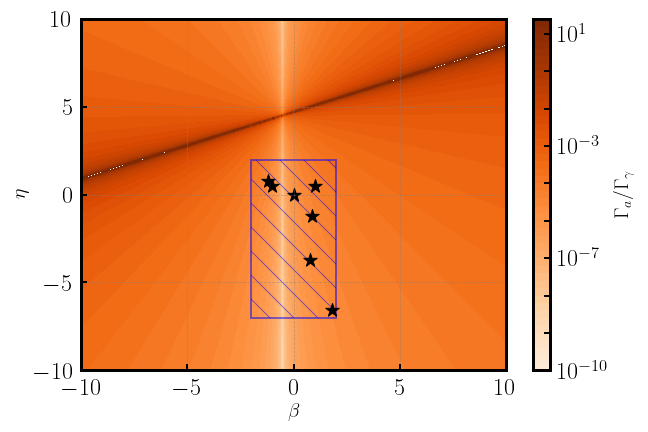}
\caption{(Color figure) Photon-axion branching as function of $\beta$ and $\eta$ for a 150-keV axion in $^{65}$Cu. The shaded area marks the parameter range covered in this work, the stars mark existing calculations, see Table.~\ref{Tab:etabeta}. Values of $\beta \sim -0.52$ minimizes the branching (gray-white area), while values along $\eta \sim 0.38\beta + 4.71$ maximize it (dark red).}
\label{fig:nucl}
\end{figure}

To estimate axion emission from a specific transition in a single isotope, the parameters $\eta$ and $\beta$ have to be known. One technique to estimate these uses a shell-model approach as done in Ref.~\cite{PhysRevD.37.618}. For magnetic dipole transitions between excited nuclear states in short lived isotopes as present in the CNO-cycle or reactors, the lack of detailed knowledge is a complication. Without detailed theoretical calculations we present a method of estimating branching using a Monte-Carlo approach. Table~\ref{Tab:etabeta} gives an overview of values used in the existing axion literature. For odd nuclei near closed shells, one can approximate that the unpaired nucleon leads to $\beta$ = $\pm$1 and $\eta$ = $\frac{1}{2}$ ~\cite{PhysRevLett.71.4120, Gavrilyuk2015}. However, for non-ground state transitions this assumption might not be valid, and the detailed calculation in Ref.~\cite{PhysRevD.37.618} shows that for off-shell nuclei the parameters can fluctuate over a wide range. We assumed input flat distributions of parameters to provide an estimate for the possible range of the photon-to-axion branching. The parameter $\beta$ was varied between -2 and +2, the parameter $\eta$ between -7 and +2. The range for both distributions were chosen so that the maximum literature values were within the distributions. The result for one specific transition can be seen in Fig.~\ref{fig:branching}. The majority of realizations fluctuate within one order of magnitude for the $\Gamma_a/\Gamma_\gamma$ branching. Outliers resulting in very low branching ratios are possible and are associated with values of $\beta \sim -0.52$ for which the numerator in Eq.~\ref{Eq:branching} approaches zero. We tested Gaussian distribution for $\eta$ and $\beta$ and only small shifts in the average value of $\Gamma_a/\Gamma_\gamma$ were found. These shifts were within the one-$\sigma$ uncertainty of the flat distribution values. Therefore, the flat distribution was used for all our calculations.
Repeating the calculation for an 1.115-MeV transition for $^{65}$Cu we found a branching of about $10^{-5.2\pm0.9}$ for an axion with mass of $m_a=$ 150\,keV, see Fig.~\ref{fig:branching}. This estimate is in good agreement with the value of $2\times 10^{-5}$ ~\cite{PhysRevD.37.618} and shows that the Monte-Carlo approach has large uncertainties but reasonable predictive power. A scan over wider ranges of the two parameters is shown in Fig.~\ref{fig:nucl}. That chosen parameter range avoid of large branching which is caused by the diminishing denominator in Eq. \ref{Eq:branching}. Therefore, the results of this study can be seen as a lower limit.

\begin{table}[]
\centering
\begin{tabular}{c | c | c | c | c}
 Isotope & transition & $\eta$ & $\beta$ & Reference \\
 \hline
$^{3}$He	 & D(p,$\gamma$)$^3$He &	0	 &	0	&~\cite{collaboration_2010}	\\
$^{7}$Li	 & 478\,keV to g.s. &	0.5	 &	1	&~\cite{collaboration_2010}\\
$^{23}$Na	 & 440\,keV to g.s. &	-1.2	&	0.88	&~\cite{PhysRevLett.66.2557}	\\
$^{55}$Mn	 & 126\,keV to g.s. &	-3.74	&	0.79	&~\cite{PhysRevLett.66.2557}	\\
$^{57}$Fe	 & 14\,keV to g.s. &	0.8 	&	-1.19	&~\cite{PhysRevLett.66.2557}	\\
$^{65}$Cu	 & 1155\,keV to g.s &	-6.59	&	1.81	&~\cite{PhysRevD.37.618}	\\
$^{83}$Kr 	& 9.4\,keV to g.s. &	0.5	 &	-1	&	\cite{Gavrilyuk2015}\\
$^{125m}$Te	 & 35\,keV to g.s. &	0.5	 &	-1	&	\cite{Derbin1997} \\
$^{139}$La 	& 166\,keV to g.s. &	0.5	 &	1	&	\cite{PhysRevLett.71.4120} \\
\end{tabular}
\caption{Tabulated values for the nuclear parameters $\eta$ and $\beta$. For $^{125m}$Te, Eq.~2 of Ref.~\cite{Derbin1997} was used and the parameters determined by comparing it to the general relation given in Eq.~\ref{Eq:1} of this manuscript.}
\label{Tab:etabeta}
\end{table}

\begin{figure}[t]
\centering
\includegraphics[width=1.0\columnwidth,keepaspectratio=true]{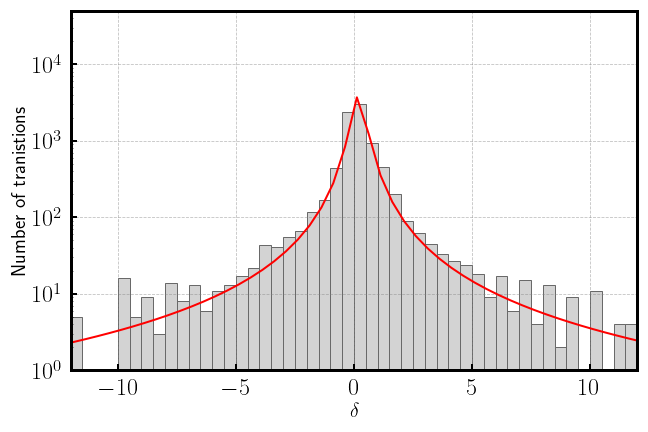}
\caption{(Color figure) Distribution of the M1/E2 mixing ratios for the magnetic dipole transitions available in the ENSDF entries of the RADWARE package. The red curve shows a fit with a Lorentzian curve. Random values using this fit were applied to transitions with unmeasured values.}
\label{fig:delta}
\end{figure}

Another uncertain parameter for a number of transitions is the M1/E2 mixing ratio $\delta$. For most stable nuclei, it is determined experimentally for low-energetic ground-state transitions. However, most of the transitions used in this study are between excited states in unstable nuclei. The majority of these have no determined M1/E2 mixing in the Evaluated Nuclear Structure Data File (ENSDF)-database. Therefore, we applied the following approach. Values were read out from the compiled data within the RADWARE package~\cite{RADFORD1995297} which uses the ENSDF data base. If a $\delta$ parameter is listed, we used it with the given uncertainty. If no uncertainty for the dipole-quadrupole mixing is available, we apply a 10\% uncertainty which corresponds to the order of uncertainty when given. For transitions with unknown $\delta$ but noted as dominantly M1, we used a Lorentzian distribution with $\Bar{\delta}= $ 0.1 and $\Gamma_\delta= $ 0.3. This curve was derived by fitting the distribution of the listed $\delta$ values of the known $M1$ transitions ($\sim10^4$), see Fig.~\ref{fig:delta}.

\section{Calculation of the axion flux}
The axion flux at a certain location can be expressed as:
\begin{equation}
\label{Eq:flux}
 \Phi_{a} = \frac{\Dot{N}}{4\pi R^2} \cdot p_{decay} \cdot p_{state} \cdot p_{branch} \cdot p_{a/\gamma} \cdot p_{travel}
\end{equation}
Here, $\Dot{N}$ describes the axion source strength, in the case of the solar calculation the number of isotopes created per time unit by solar fusion. Similarly, $\Dot{N}$ describes in the reactor environment the number of short lived isotopes created in the fusion process. For the solid angle factor $1/4\pi R^2$ we assume a point-like isotropic emission of axions, where $R$ is the distance between axion source and detector. 

\begin{figure*}[t]
\centering
\includegraphics[width=2.0\columnwidth,keepaspectratio=true]{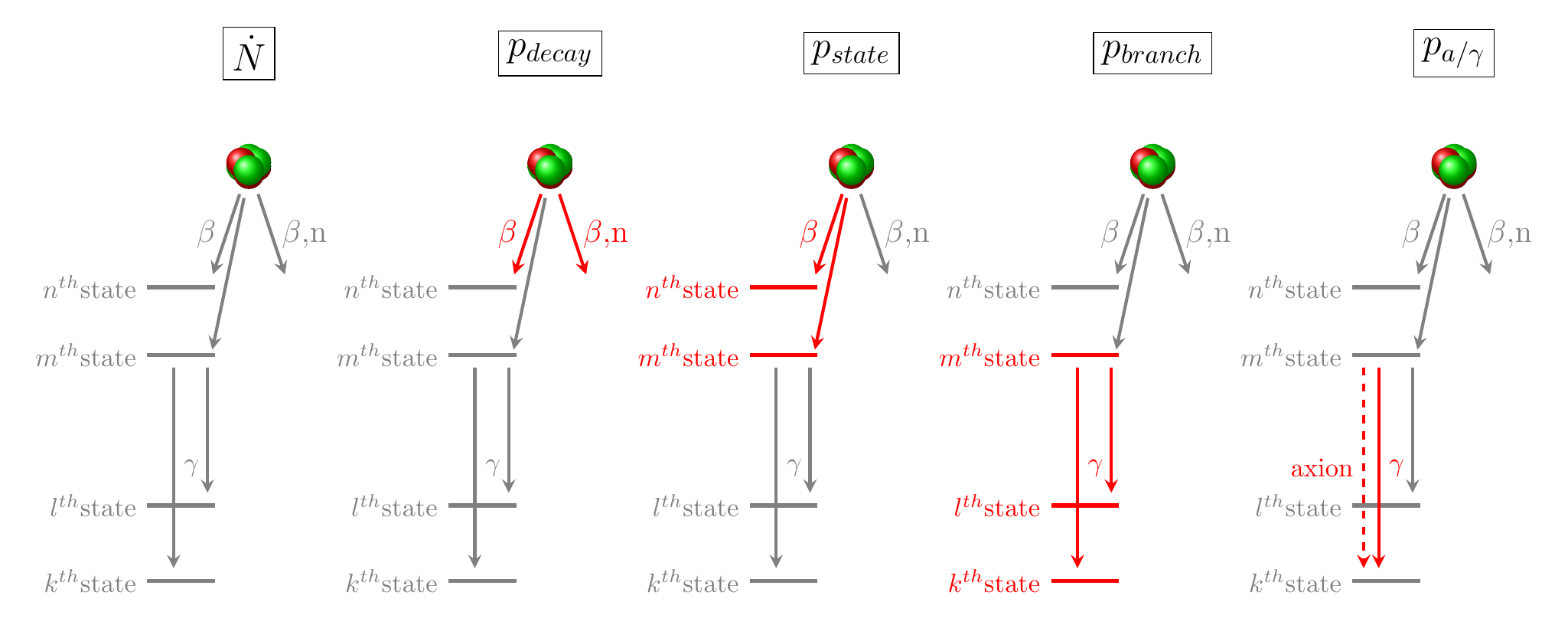}
\caption{(Color figure) Schematic illustration of Eq.~\ref{Eq:flux} for the case of an isotope that could undergo a $\beta$-decay or a $\beta,n$-decay. The different factors are highlighted in red. For each of the five steps, the red highlighted part depicts the factor above.}
\label{fig:decayscheme}
\end{figure*}

The $p$ factors describe various probabilities for the creation of states that can create axions, the axion emission, as well as the geometric probability and are illustrated in Fig.~\ref{fig:decayscheme}. The factor $p_{decay}$ describes the probability for an isotope to be created and undergo a certain decay. This factor is equal to one for most decays for the solar environment since only one decay is possible. In the reactor calculation it describes the probability for an isotope to be created by a uranium fission as well as the branching of a ($\beta$)-decay, a ($\beta,n$)-decay or a other decay processes through which the short lived isotopes decay. The factor $p_{state}$ indicates the probability that a specific excited state is populated in the daughter nucleus following a radioactive decay. As shown in the middle of Fig.~\ref{fig:decayscheme}, the number $p_{branch}$ takes into account the decay branching of the individual excited states. Often the states created deexcite through a cascade of $\gamma$ rays. The axion-to-photon probability $p_{a/\gamma} = \frac{\Gamma_a}{\Gamma_\gamma}$ describes for one of these individual transitions how probable it is for an axion to be created. As discussed in Sec.~\ref{sec:axionphoton}, it can be expressed as the ratio of the two widths. Finally, the last factor, $p_{travel}$ takes into account that axions can decay while traveling from the source to the detector. This factor will be discussed in the next section.

\section{\label{sec:axiontime}Axion lifetime}
\begin{figure}[t]
\centering
\includegraphics[width=1.0\columnwidth,keepaspectratio=true]{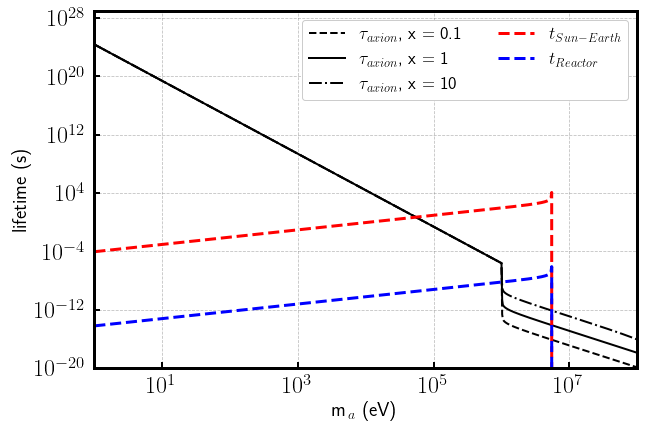}
\caption{(Color figure) Lifetime for a 5.5-MeV axion as created in the solar burning process. The lifetime of the standard QCD axion (black, different values of $x$) decreases at 1.022\,MeV where the electron-positron pair production channel opens up. The flight time to Earth (red) prevents the observation of high-mass axions at the Earth, even when relativistic effects are included. In addition, the flight times for a reactor based experiment using a 10-m long flight path is shown as well (blue).}
\label{fig:lifetime}
\end{figure}

The decay of solar axions in flight can significantly reduce the sensitivity of Earth-bound searches. The decay width of keV-MeV mass axions consists of two major channels, the decay into photons, $\Gamma_{a\rightarrow{}\gamma\gamma}$ and the decay into an electron-positron pair, $\Gamma_{a\rightarrow{}e^+e^-}$. The latter only occurs if the mass of the axion is larger than 1.022\,MeV. Therefore, every Earth-bound detector is expected to see only a part of the total flux. Similar corrections were applied in previous studies~\cite{PhysRevD.38.3375}. We follow the description of the axion decay widths as given in Ref.~\cite{Altmann1995}. 
\begin{equation}
 \Gamma(a\rightarrow{}\gamma\gamma) = z^{-1} \Big( \frac{m_a}{m_\pi} \Big)^5 \Gamma(\pi^0\rightarrow{}2\gamma)
\end{equation}
\begin{equation}
 \Gamma(a\rightarrow{}e^+e^-) = \frac{m_e \sqrt{m_a^2 - 4m_e^2}}{8\pi f_a^2 x^2}
\end{equation}

\begin{equation}
 \tau = \frac{1}{\Gamma_{total}} =\frac{1}{\Gamma(a\rightarrow{}\gamma\gamma) + \Gamma(a\rightarrow{}e^+e^-)}
\end{equation}
The decay width into two photons is proportional to the pion decay width $\Gamma(\pi^0\rightarrow{}2\gamma)$, given as 7.82\,eV~\cite{PhysRevLett.106.162303}. The second decay channel not only depends on the electron mass but also includes the vacuum expectation value $x$ of the two Higgs-doublet~\cite{PhysRevD.18.1607} to which the electron is coupled. This value is usually estimated to be one~\cite{Altmann1995}. Fig.~\ref{fig:lifetime} shows the effect of the life time for different values of $x$.

As shown in Fig.~\ref{fig:lifetime} the lifetime for the QCD axion greatly decreases for masses above the threshold for pair production. This makes an observation of higher mass axions from the Sun impossible since the travel time to Earth is at least 8.3 minutes. Time dilating effects are included and extend the observable mass range by around a factor 2. The probability to see an axion can be calculated as:
\begin{equation}
p_{travel} = e^{(-\frac{t}{\tau}) },
\end{equation}
where $t$ is the travel time for a certain distance. The velocity of the axion can be calculated since the total energy corresponds to the energy of the transition $E_\gamma$. As Fig.~\ref{fig:lifetime} shows, even experiments with an axion source near the detector suffer lifetime constraints and create an experimental challenge for the direct detection of heavy axions.

\section{Solar axions revisited}
The Sun produces energy by a number of processes, in particular fusion of hydrogen and helium, as well as several proton capture reactions. For capture reactions, the number, energy, and multipole character of photon emission varies from reaction to reaction. The cross section can be split into two parts, a direct capture process, and a capture into resonant states, which can lay below or above the separation threshold. In the direct capture process, e.g., $X+p\rightarrow X^{\prime} + \gamma$ where $X$ is the target nucleus, $p$ is the incoming proton, $X^{\prime}$ is the new formed nucleus and $\gamma$ is the outgoing photon, mostly the ground state of the newly formed nucleus is populated and the photon is emitted with an energy given by the total reaction energy, $E_\gamma = S_p + E_p$ where $S_p$ is the sum of the proton separation energy of the nucleus $X^\prime$ and $E_p$ is the incoming proton energy. In some cases the population of the first excited state of $X^\prime$ is possible and the energy of the photon is reduced by this amount $E_\gamma = S_p + E_p - E_1$. 

For resonant capture, the peak energy is $E_x=S_p+E_p$. These resonances have a certain width and can extend above or below the separation threshold. This can cause a sub-threshold resonance significantly contributing to the total cross section in which the decay cascades of the excited states are included. Several of these patterns contain magnetic dipole transitions. 

Experimental data at solar energies for direct capture is sparse due to the low rates in Earth-bound laboratories. A number of resonances have been studied at higher proton energies and their contribution in a solar energy regime can be extrapolated. In general, we follow the discussion and data given in Refs.~\cite{HUANG2010824,XU201361} to estimate the contributions of the individual parts. For specific reactions, we also cite the individual literature containing specific information. 

\begin{itemize}
 \item \textbf{$^{3}$He production}\\
 Low-background underground detectors looking for neutrinoless double beta decay or dark matter can be used to search for such a signal. This reaction as an axion source has been studied before by several groups~\cite{RAFFELT1982323, collaboration_2010,PhysRevD.85.092003, Derbin2013}. The probability to emit a photon with magnetic dipole properties lies at around 43\% and there is no mixture of an electric quadruple assumed~\cite{collaboration_2010}. The transition energy is very high at 5.5\,MeV which has advantages since most background from natural radioactivity occurs at lower energies than the expected signal. \\
 \item \textbf{$^7$Be-production} The production of short-lived $^7$Be proceeds through two channels. For the resonant capture, the proton energy $E_p$ needs to be high enough to populate higher excited states. The first available state is at 6.73\,MeV and requires a minimum 1.1\,MeV kinetic energy of the proton. For the kinetic energies of the protons in the Sun's core we assume a Boltzman distribution with a temperature of 1.3\,keV (corresponding to the core temperature 1.5$\times$10$^7$ K). Therefore, this reaction is highly suppressed ($p_{state}\sim0$). The direct capture channel results in the population of effectively two states, the ground state and the first excited state in $^7$Be~\cite{HUANG2010824}. Photons produced in this direct capture have energies of 5.1 and 5.5\,MeV and are dominantly E1-nature, for which we assume a negligible M1 contribution ($p_{branch}\ll 1$). However, when produced the first excited state of $^7$Be decays via an M1-transition to the ground state for which we can calculate an axion mixture. In order to calculate the rate, we assume nearly all created excited state beryllium atoms decay to the ground state and then undergo $\beta$-decay. Hence, we use this rate given by neutrino flux calculations. \\
 \item \textbf{$^7$Be-decay} \\
 Isotopes of $^7$Be decay to $^7$Li via electron capture to an excited state at 471\,keV 10\% of the time~\cite{SZABO1972527}. This state's deexcitation is dominated by a magnetic dipole transition. This reaction rate can be directly connected to the reaction rate calculated by Bahcall~\cite{Bahcall_2001}. 
 \item \textbf{$^8$B-production} \\
 The proton-separation energy $S_p$ of $^8$B is only 137\,keV. The first excited state that could decay via an M1-transition is located at 774\,keV. Again, the temperature of the Sun completely suppresses the resonant capture to the first excited state and we set $p_{state}$ to zero. Instead a direct capture process takes place that populates the ground-state of $^{8}$B while emitting a $\gamma$-ray with energy $E_\gamma = S_p + E_p \sim S_p$. In the case of $^8$B, only one sub-threshold state exists. For available proton energies s-wave proton capture is assumed to dominate the reaction rate, resulting in a dominant E1-nature of the emitted photon~\cite{PhysRevC.35.363, PhysRevC.60.015803}. From these references, we used $p_{branch}<$ 10$^{-3}$ as an upper limit for the M1-contribution.
 \item \textbf{$^{13}$N-production} \\
 The CNO-cycle contributes a few percent of the stellar energy production~\cite{PhysRev.55.434} and burns about 1\% of the solar hydrogen~\cite{Wiescher2010}. The first reaction that we discuss in this cycle is the proton capture on carbon. Again the reaction can proceed through direct and resonant capture. Resonant reactions have promising magnetic dipole transitions, e.g. at 2.3\,MeV, and contribute to about 70\% to the cross section~\cite{HUANG2010824}. The direct reaction channel is dominated by E1-transitions~\cite{PhysRevC.64.065804} and for the calculation we estimate that the M1 contribution is in the range of the contributions discussed for the $^8$B-production. The reaction rate for this capture can be estimated by using the well calculated neutrino rate of the following decay under the assumption that all produced $^{13}$N undergo decay and are not consumed by other reaction channels. The decay itself is a pure electron-capture process that has no M1-deexcitations. 
 \item \textbf{$^{14}$N-production} Similar to the production of $^{13}$N, the resonant capture dominates the capture process cross section for solar energies. The tails of the first resonant state contribute to the S-factor at solar energies by about 70\%~\cite{HUANG2010824,GUOBing}. The majority of this contribution is coming from the second resonance above the proton threshold, and not from the very narrow first resonance~\cite{MUKHAMEDZHANOV2003279}. The nature of the direct capture is similar to the previous cases assumed to be electric dipole with only negligible contributions by other multipoles.
 \item \textbf{$^{15}$O-production} In a solar environment, this reaction is the slowest process within the CNO cycle. Several resonances add up to the total cross section while the direct capture to the ground state only contributes only about 15\%~\cite{FORMICOLA200461}. This direct capture is dominantly of E1 nature. The resonant capture populates a number of states, of which only the 6859-keV state consists of a M1-transition. As shown in Ref.~\cite{FORMICOLA200461} this resonance does not contribute to the cross section at the energy range of interest. 
 The neutrino flux of the following decay is very well studied~\cite{Bahcall_2001}. The expected neutrino flux on Earth is slightly lower than the rate of the initial $^{13}$N decay and the geometrical factor, due to the opacity of the Sun. We assume that the axions are not effected, and use the solar reaction rate. 
 \item \textbf{$^{16}$O-production} This first CNO-breakout reaction~\cite{Wiescher_1999} is responsible for the creation of the oxygen content of the Sun since $^{15}$O has a short half-life. The reaction rate of this production can be estimated based on the neutrino flux of the follow-up production and decay of $^{17}$F. Here, we assume that at least as much oxygen is produced as is consumed in the production of fluorine ($^{17}$F). The reaction cross section is a sum of contributions by the first few resonances above the proton binding energy and a direct capture component. As discussed in Ref.\,\cite{ROLFS1974450}, about 60\% of the contribution to the S-factor, which is the factor separating out the Coulomb interaction (barrier) energy term from the cross-section, at solar energies are due to direct capture. 
 \item \textbf{$^{17}$F-production} The $^{17}$F nucleus has a relatively low proton separation threshold of only 600\,keV. The production process is dominated by direct capture to the first excited state which has a small magnetic dipole contribution~\cite{HUANG2010824, BENNACEUR200075}. The extended CNO cycle is closed by the $^{17}$O$(p,\alpha)^{14}$N reaction. The produced fluorine is either consumed by this reaction or the decay to oxygen. Neither contains a magnetic dipole transition. But studied within the solar neutrino problem, the decay rate can be used for the production rate assuming $^{17}$F production and decay are in, or close to, equilibrium.
 
\end{itemize}

\subsection{Discussion}
\begin{figure}[t]
\centering
\includegraphics[width=1.0\columnwidth,keepaspectratio=true]{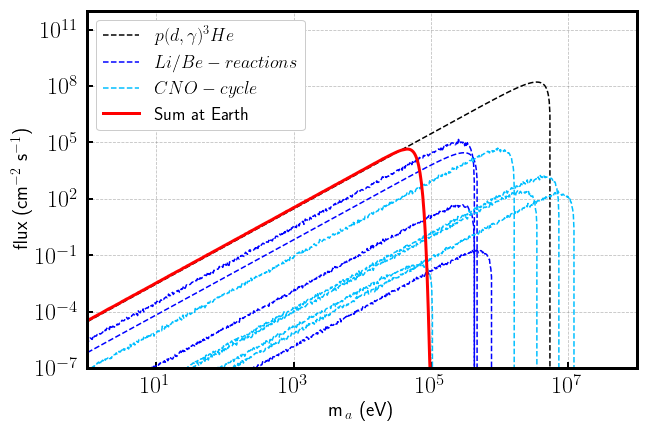}
\caption{(Color figure) Expected axion flux from the Sun by an axionic component within magnetic dipole transitions. The dashed lines show which spectrum would be visible by the individual contributors in the solar environment without decay-in-flight. The red solid line represents the sum and takes into account that axions decay on the way to the Earth. }
\label{fig:fluxSun}
\end{figure}

\begin{figure}[t]
\centering
\includegraphics[width=1.0\columnwidth,keepaspectratio=true]{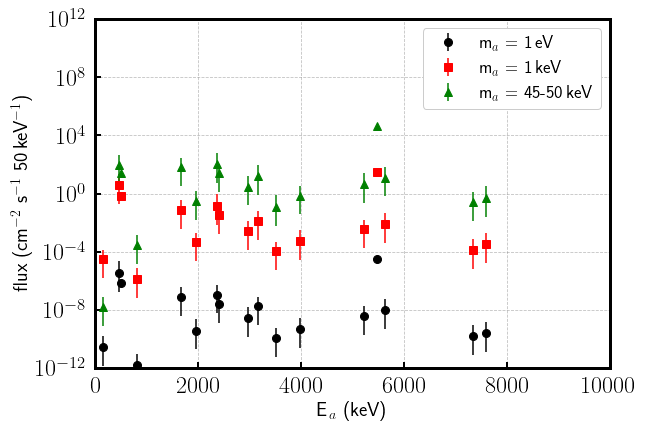}
\caption{(Color figure) Total flux of the solar axions for different axion masses. For an axion mass of $\sim$45\,keV the flux, shown in Fig.~\ref{fig:fluxSun}, reaches its maximum for a detector situated at the Earth radius distance.}
\label{fig:energySpecSun}
\end{figure}

Figure~\ref{fig:fluxSun} shows that the dominating contribution to a solar axion flux using Eq.~\ref{Eq:flux} would come from the proton-deuterium reaction. The figure also shows that the sensitivity increases with increasing axion mass. When the axion mass approaches the transition energy, a sharp turn-off can be observed, due to the $E_\gamma^2-m_a^2$ factor in Eq.~\ref{Eq:branching}. The solar maximum of the emitted flux is at $m_{a}=3.5$\,MeV, however the maximum energy when arriving at Earth is expected to be between 45 and 50\,keV. As discussed, the axion lifetime plays a significant role. Along the long flight path from the Sun to the Earth massive axions will decay. This effect limits the observation of massive solar axions and requires alternative research approaches. The uncertainties of the total flux due to unknown nuclear parameters are small since the parameters for the strongest contributor are known, as shown in Table~\ref{Tab:solarisotopes}. Figure~\ref{fig:energySpecSun} shows the expected axion-energy spectrum for different axion mass values. The sensitive axion mass is restricted to be less than 100\,keV due to the decay in flight. However, the total observable energy in a detector can be higher since it corresponds to the energy of the original transition. As shown, a number of these are higher than the natural background radiation coming e.g. from $^{208}$Tl decays. In contrast to previous searches, this approach predicts a spectrum of correlated peaks that arises from various contributors. The search for multiple correlated peaks should improve the overall sensitivity, and can serve as internal cross-check.

\section{Axions from Nuclear Reactors}
A number of isotopes contribute to the power production of a reactor. For this study we focus on the $^{235}$U which is the dominant source of creating fission isotopes. On average, the neutron-induced fission of one nucleus $^{235}$U releases about 200\,MeV of energy~\cite{PhysRev.55.511.2}. Additional isotopes can be created when fissioning other uranium or even plutonium isotopes. Therefore, the calculation presented here can be seen as a lower limit. According to ENDF/B-VII~\cite{CHADWICK20112887} about 190\,MeV of the total energy is contained inside a reactor, the remaining being carried away by neutrinos. It can be estimated that a 85-MW$_{thermal}$ reactor like HFIR (High Flux Isotope Reactor) which is highly enriched in $^{235}$U undergoes about 2.8$\times$10$^{18}$ fissions per second~\cite{doi:10.1080/00295639.2017.1292090}. Detailed simulations~\cite{ORNL1, ORNL2} show neutron production rates of $>$10$^{18}$ neutrons per second in the core, which is in agreement with the estimate given that one fission process creates multiple neutrons. 

An overview on calculations and measurements of isotopes created in the fission process is given in Ref.~\cite{PhysRevC.58.905}. Following the same approach as for the solar production we analyzed where magnetic dipole transitions occur in the decay scheme by using the ENSDF database. Only transitions for which the database has assigned a magnetic dipole type were taken into account. By including the decay chains of fission products, we found 390 magnetic dipole candidate transitions for the isotopes given by Ref.~\cite{PhysRevC.58.905}. This number can be seen as a lower limit since data for the short lived isotopes are sparse and the multi-polarities are often not determined. However, strong transitions were likely already identified, measured, and are therefore, listed in the database. So it is safe to assume that potential contributions from unknown transitions are small.

\subsection{Discussion} 
\begin{figure}[t]
\centering
\includegraphics[width=1.0\columnwidth,keepaspectratio=true]{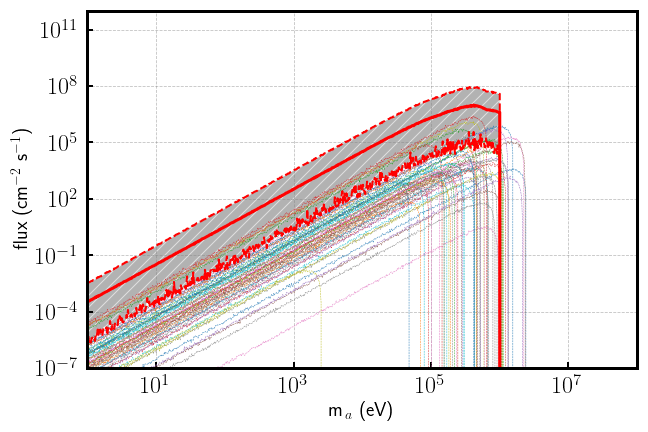}
\caption{(Color figure) Expected axion flux from HFIR (85 MW$_{th}$) at a 10-meter distance due to a possible mixing of an axionic component into the deexcitation of magnetic dipole transitions. The red solid line represents the sum of the individual contributions (colored). The dashed lines enclose the area between the worst and best case of all Monte-Carlo realizations. The short half-lives of heavy axions limit the observable masses to less than 1.022\,MeV.}
\label{fig:fluxReactor}
\end{figure}

\begin{figure}[t]
\centering
\includegraphics[width=1.0\columnwidth,keepaspectratio=true]{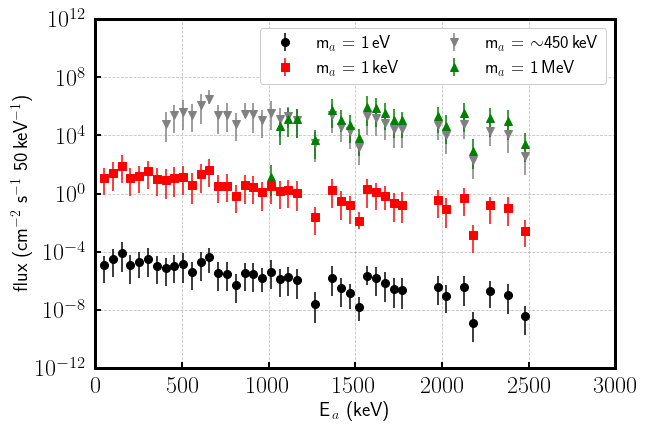}
\caption{(Color figure) Total energy spectrum of the reactor axions for different axion masses. The same experimental condition as in Fig.~\ref{fig:fluxReactor} are used for this spectrum. For an axion mass of the $\sim$450\,keV the flux reaches its maximum.}
\label{fig:energySpecReactor}
\end{figure}

As Table~\ref{Tab:reactorisotopes} shows, hundreds of possible magnetic dipole transition will contribute to a possible axion flux from an active reactor. Most of these occur between excited states in short lived isotopes. The parameters $\beta$ and $\eta$ are not calculated yet for the majority of these transitions. As described, we used a Monte-Carlo approach and calculated 10000 possible combinations of nuclear parameters for each isotope. Figure~\ref{fig:fluxReactor} shows the mean value of each isotope contribution. In contrast to the solar flux, no dominant contribution by transition can be found. The maximum sensitivity is given for an axion mass of about 450\,keV. This is slightly lower than the for the solar axion spectrum at emission. However, due to the short flight path of the axions, a reactor-based experiment does not suffer as heavily from in-flight decays. Therefore, it can be sensitive to axion masses up to 1\,MeV, cf. Fig.~\ref{fig:lifetime}. For a detector placed 10-meters from the HFIR core (85 MW$_{th}$), the axion flux would be within an order of magnitude of that from the Sun for light axions.
Figure~\ref{fig:energySpecReactor} shows the energy spectrum at a reactor. Comparing the results from Fig.~\ref{fig:energySpecSun} with the reactor spectrum the following points can be made. While the solar spectrum is dominated by a few individual transitions, the reactor spectrum is a compilation of numerous transitions over a broad range of energies. The reactor spectrum has a lower endpoint than the solar spectrum. As Fig.~\ref{fig:energySpecReactor} indicates, the spectrum can have a lower endpoint for particular axion masses. If $E_{\gamma} < m_{a}$ no axion can be produced, hence a minimum transition energy is necessary to create an axion with a certain mass. Therefore, the existence or non-existence of a lower endpoint can help to restrict the axion mass further. This feature can not be observed in the solar axion flux since the in-flight decay restricts the observable axion mass range to masses below the lowest transition energy.

\begin{figure}[t]
\centering
\includegraphics[width=1.0\columnwidth,keepaspectratio=true]{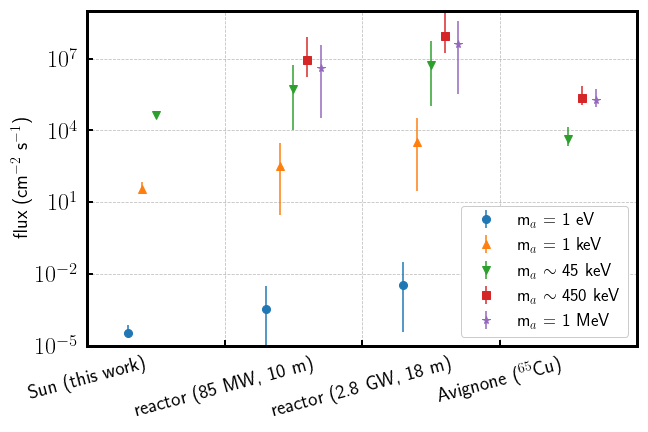}
\caption{(Color figure) Total axion flux from different sources. For reactor based searches, we choose conditions as possible at HFIR, or a power reactor~\cite{Altmann1995}. The expected axion flux from Ref.~\cite{PhysRevD.37.618} is calculated using the branching from Fig.~\ref{fig:branching}, a 10.9\,kCi-strong $^{65}$Cu source, and a detector at 20\,inches distance.}
\label{fig:results}
\end{figure}

\section{Detection of axions}
\begin{figure}[t]
\centering
\includegraphics[width=1.0\columnwidth,keepaspectratio=true]{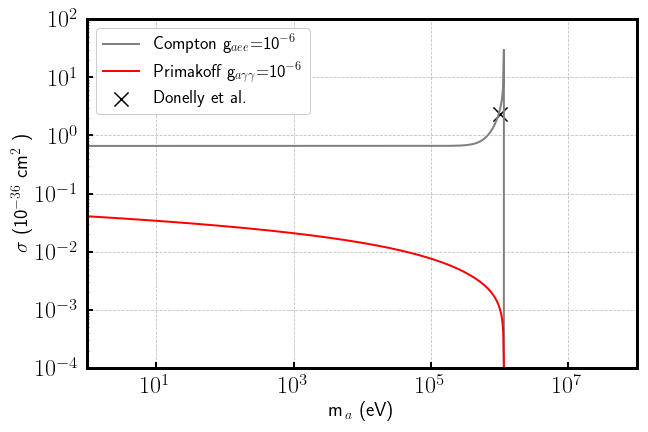}
\caption{(Color figure) Cross section for Primakoff- (red), and Compton-conversion (black) as a function for axion mass for a 1.115-MeV transition as in Fig.~\ref{fig:branching}. For both cross sections the coupling constants $g_{a\gamma\gamma}$ and $g_{aee}$ are set to 10$^{-6}$. The cross marks the prediction from Ref.\,\cite{PhysRevD.18.1607}.}
\label{fig:crosssection}
\end{figure}

\begin{figure}[t]
\centering
\includegraphics[width=1.0\columnwidth,keepaspectratio=true]{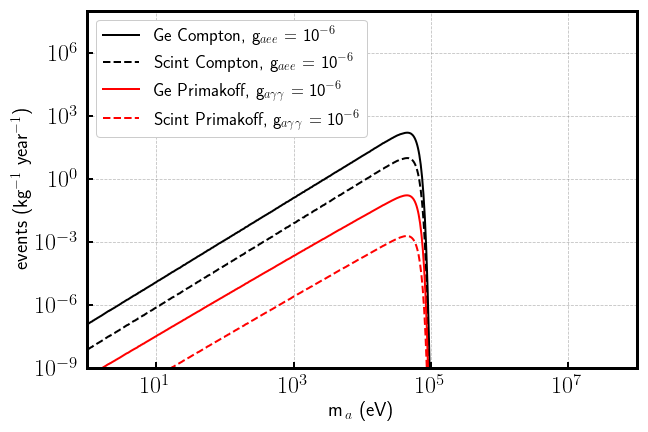}
\includegraphics[width=1.0\columnwidth,keepaspectratio=true]{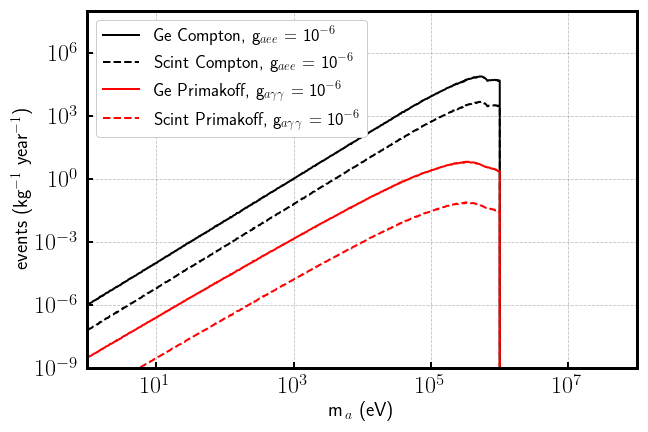}
\caption{(Color figure) Expected integral count rate for the axion spectra shown in Fig.~\ref{fig:energySpecSun} and~\ref{fig:energySpecReactor} for one kilogram of germanium or carbon-based scintillation detector. The flux value means are given for solar axions on the Earth (top figure) and a HFIR-reactor based experiment (bottom), cf. Figs.~\ref{fig:fluxSun} and~\ref{fig:fluxReactor}. Count rates were calculated with the $g_{a\gamma\gamma}$ and $g_{aeee}$ = 10$^{-6}$ as input.}
\label{fig:countrate}
\end{figure}

\begin{figure}[t]
\centering
\includegraphics[width=1.0\columnwidth,keepaspectratio=true]{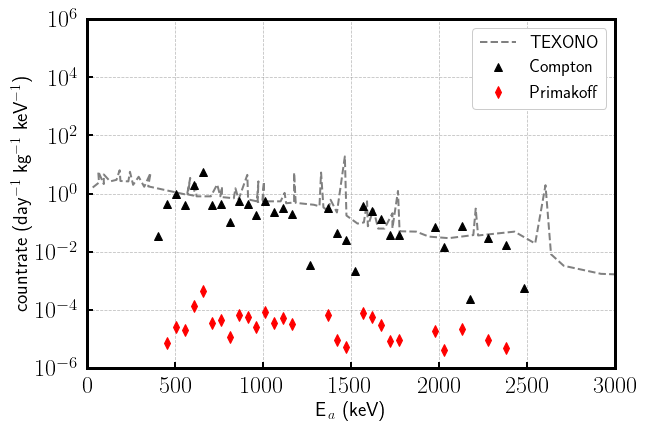}
\caption{(Color figure) Count rate over 50-keV bins as calculated from the spectrum in Fig.~\ref{fig:energySpecReactor} for an axion mass around 450 keV. The two detection processes, Primakoff (green) and Compton (red), result in different detection count rates using the current limits. The dashed line shows the achieved background from Ref.~\cite{PhysRevD.75.052004}.}
\label{fig:countrate2}
\end{figure}

The detection of axions is assumed to use two dominant process~\cite{PhysRevD.37.618, PhysRevD.75.052004, PhysRevD.85.092003}. The Primakoff and the Compton conversion coupling are considered here. An axion flux, $\Phi_a$, would create a count-rate $S$ as,
\begin{equation}
 S_{Compton} = \Phi_a Z n \sigma_{C}\, \epsilon
\end{equation}
\begin{equation}
 S_{Primakoff} = \Phi_a n \sigma_{P}\, \epsilon
\end{equation}
Here, $n$ represents the number of atoms per unit detector mass, $Z$ stands for the number of electrons per atom in the Compton process. The detection efficiency $\epsilon$ of the created electron or photon is assumed to be unity for this study. The two cross sections $\sigma_{C}$ and $\sigma_{P}$ are used as defined in Ref.~\cite{PhysRevD.37.618} with the coupling constants defined as in Ref.~\cite{PhysRevD.85.092003}:
\begin{equation}
\begin{aligned}
 \sigma_C = & g_{aee}^2 \frac{\alpha}{8 m_e^2 k_a} \Bigg[\frac{2m_e^2(m_e + E_a)(2m_e E_a + m_a^2)}{(m_e^2 + (2m_e E_a + m_a^2))^2} \\
 & + \frac{4 m_e(m_a^4 + 2 m_a^2 m_e^2 - 4 m_e^2 E_a^2)}{(2m_e E_a + m_a^2)(m_e^2 + (2m_e E_a + m_a^2) )}\\
 & + \frac{4 m_e^2 k_a^2 + m_a^4}{k_a (2m_e E_a + m_a^2)} \ln{\frac{m_e + E_a + k_a}{m_e + E_a - k_a}} \Bigg] \\
\end{aligned}
3\end{equation}
\begin{equation}
 \sigma_P = g_{a\gamma\gamma}^2 \frac{Z^2 \alpha}{2 \beta_k} \Big[ \frac{1+ \beta_k^2}{2 \beta_k} \ln{\frac{1+ \beta_k}{1- \beta_k}} -1 \Big]
\end{equation}
The factor $ \beta_k$ is given as the ratio of $k_a$ and the $k_\gamma$. For a 1.115-MeV transition the cross section in Germanium is shown in Fig.~\ref{fig:crosssection}. For this calculation the current best limits for the two coupling constants in the high-keV axion mass range as given in Ref.~\cite{PhysRevD.75.052004} are used. For two different detector types, a Ge detector and a carbon-based scintillation detector, the expected count rates are shown in Fig.~\ref{fig:countrate}. The high flux of axions at a reactor results in up to 10$^5$ counts/(kg yr) over a broad energy range. Current results of the TEXONO collaboration~\cite{PhysRevD.75.052004} achieved competitive backgrounds in a reactor environment, cf. Fig.~\ref{fig:countrate2}. The figure also shows that a spectral search has an advantage over the search for individual signatures since individual lines can be suppressed. 

\section{Summary}
In this manuscript we show that the search for axions or axion-like particles should be extended from a peak-above background search to a spectral analysis. This extension uses the same model and physics as peak based searches in a complex environment. 
In contrast to single peak searches, we expect that the search for a spectral excess opens the axion field beyond ultra low-background, low-threshold detectors. We showed that the basic approach assuming that axions are produced in magnetic dipole transitions can be extended from the Sun as the axion source. Nuclear reactors, in particular research reactors allow a search over a wider mass range. The decay of the axion remains an experimental challenge for masses greater than 1\,MeV that can not be solved easily. Using the current limits on the coupling factors for ALPs in the\,keV to\,MeV range, we estimated that data from current low-background experiments should be sufficient to improve the limits. Under the assumption that the background rates of next generation experiments can be improved, reactor-based experiments should be able to compete with limits from astrophysical observations and models.

\begin{acknowledgements}
This material is based upon work supported by the U.S. Department of Energy, Office of Science, Office of Nuclear Physics under Award Number LANL-E9BW/EM77/EM78. We gratefully acknowledge the support of the U.S.~Department of Energy through the Los Alamos National Laboratory LDRD Program for this work. We would also like to thank R. Henning and D. Alves for many discussions and references.
\end{acknowledgements}
\FloatBarrier
\bibliography{main.bib}

\newpage

\appendix*

\begin{table*}[]
\centering
\begin{tabular}{c | c c c c c | c c c | c}
 reaction & $E_\gamma$ (keV) & $\Dot{N}$ (s$^{-1}$) & $p_{decay}$ & $p_{state}$ & $p_{branch}$ & $\delta$ & $\beta$ & $\eta$ & comment\\
 \hline 
 & & & & & & & & &\\
 $p(d,\gamma)^3He$ & 5439 & 1.6$\times$10$^{38}$ & 1 & 1 & 0.43 & 0 & 0 & 1 &\\
 & & & & & & & & &\\
 $^{6}$Li(p,$\gamma$)$^{7}$Be (direct) & 5605 & 1.3$\times$10$^{37}$ & 1 & 0.7 & $<$ 10$^{-3}$ & 0 &n.a. & n.a. &\\
 $^{6}$Li(p,$\gamma$)$^{7}$Be$^*$ (direct) & 5176 & 1.3$\times$10$^{37}$ & 1 & 0.3 & $<$ 10$^{-3}$ & 0 &n.a. & n.a. & to 1$^{st}$ excited state\\
 $^{6}$Li(p,$\gamma$)$^{7}$Be (resonant) & 429 & 1.3$\times$10$^{37}$ & 1 & $<$ 10$^{-4}$ & 1 & 0 &n.a. & n.a. &\\
 $^{7}$Be$^*$ $\xrightarrow{}$ $^{7}$Be & 429 & 1.3$\times$10$^{37}$ & 1 & 0.3 & 1 & 0 &n.a. & n.a. & decay of the 1$^{st}$ excited state\\
 & & & & & & & & &\\
 $^7$Be$\xrightarrow{}$ $^7$Li + $\nu$ + $\gamma$ & 477 & 1.3$\times$10$^{37}$ & 1 & 1 & 0.1 & 1 & 0.5 & 0.2 &\\ 
 & & & & & & & & &\\
 $^{7}$Be(p,$\gamma$)$^{8}$B (direct) & 137 & 1.5$\times$10$^{34}$ & 1 & 1 & $<$ 10$^{-3}$ & 0 &n.a. & n.a. &\\
 $^{7}$Be(p,$\gamma$)$^{8}$B (resonant) & 774 & 1.5$\times$10$^{34}$ & 1 & $<$ 10$^{-4}$ & 1 & 0 &n.a. & n.a. &\\
 & & & & & & & & &\\
 CNO reactions & & & & & & & & &\\
 & & & & & & & & &\\
 $^{12}$C(p,$\gamma$)$^{13}$N (direct) & 1943 & 1.3$\times$10$^{36}$ & 1 & 0.3 & $<$ 10$^{-3}$ & 0 &n.a. & n.a. &\\
 $^{12}$C(p,$\gamma$)$^{13}$N (resonant) & 2364 & 1.3$\times$10$^{36}$ & 1 & 0.7 & 0 & 0&n.a. & n.a. & to 1$^{st}$ resonance above $S_p$\\
 $^{12}$C(p,$\gamma$)$^{13}$N (resonant) & 3502 & 1.3$\times$10$^{36}$ & 1 & $<$ 10$^{-4}$ & 1 & 0 &n.a. & n.a. & to 2$^{nd}$ resonance above $S_p$\\
 & & & & & & & & &\\
 $^{13}$C(p,$\gamma$)$^{14}$N (direct) & 7550 & 1.3$\times$10$^{36}$ & 1 & 0.3 & $<$ 10$^{-3}$ & 0 &n.a. & n.a. & \\
 $^{13}$C(p,$\gamma$)$^{14}$N$^*$ (resonant) & 3948 & 1.3$\times$10$^{36}$ & 1 & 0.7 & 0.005 & 2.8 &n.a. & n.a. & decay of the 7966-keV state\\
 $^{13}$C(p,$\gamma$)$^{14}$N$^*$ (resonant) & 3148 & 1.3$\times$10$^{36}$ & 1 & 0.7 & 0.018 & 0 &n.a. & n.a. & decay of the 7966-keV state\\
 $^{13}$C(p,$\gamma$)$^{14}$N$^*$ (resonant) & 2956 & 1.3$\times$10$^{36}$ & 1 & 0.7 & 0.003 & 0 &n.a. & n.a. & decay of the 7966-keV state\\
 $^{13}$C(p,$\gamma$)$^{14}$N$^*$ (resonant) & 2371 & 1.3$\times$10$^{36}$ & 1 & 0.7 & 0.035 & 0 &n.a. & n.a. & decay of the 7966-keV state\\
 $^{13}$C(p,$\gamma$)$^{14}$N$^*$ (resonant) & 2312 & 1.3$\times$10$^{36}$ & 1 & 0.7 & 0.159 & 0 &n.a. & n.a. & decay of the 7966-keV state\\
 $^{13}$C(p,$\gamma$)$^{14}$N$^*$ (resonant) & 1635 & 1.3$\times$10$^{36}$ & 1 & 0.7 & 0.121 & 0 &n.a. & n.a. & decay of the 7966-keV state\\
 & & & & & & & & &\\
 $^{14}$N(p,$\gamma$)$^{15}$O (direct) & 7296 & 1.3$\times$10$^{36}$ & 1 & 0.15 & $<$ 10$^{-3}$ & 0 &n.a. & n.a. & \\
 & & & & & & & & &\\
 extended CNO & & & & & & & & &\\
 & & & & & & & & &\\
 $^{15}$N(p,$\gamma$)$^{16}$O (direct) & 12127 & 1.5$\times$10$^{34}$ & 1 & 0.6 & $<$ 10$^{-3}$ & 0 &n.a. & n.a. & \\
 & & & & & & & & &\\
 $^{16}$O(p,$\gamma$)$^{17}$F (direct) & 105 & 1.5$\times$10$^{34}$ & 1 & 1 & $<$ 10$^{-3}$ & 0 &n.a. & n.a. & to 1$^{st}$ excited state\\
\end{tabular}
\caption{Parameters for the calculation of the individual axion-branching and flux contributions. For transitions with unknown nuclear parameters ($n.a.$) distributions of parameters were applied as discussed in Sec.~\ref{sec:parameters}.}
\label{Tab:solarisotopes}
\end{table*}

\clearpage

\begin{center}
\begin{longtable}{c | c c c c | c }

reaction & $E_\gamma$ & $p_{decay}$ & $p_{decay}^*$ & $p_{branch}$ & $\delta$ \\
 & (keV) & [$\%$] & [$\%$] & & \\
\hline
\endfirsthead

reaction & $E_\gamma$ & $p_{decay}$ & $p_{decay}^*$ & $p_{branch}$ & $\delta$ \\
 & (keV) & [$\%$] & [$\%$] & & \\
\hline
\endhead

\hline
\endfoot

\hline
\caption{Parameters for the calculation of the individual axion-branching and flux contributions. For transitions with unknown nuclear parameters ($n.a.$) distributions of parameters were applied as discussed in Sec.~\ref{sec:parameters}. The quantity $p_{decay}$ refers to the values given in Ref.~\cite{PhysRevC.58.905}, while $p_{decay}^*$ includes feeding of daughter decays as well and branching of decays into $\beta$ and $\beta \, n$-decays. The factors $p_{state}$ and $p_{branch}$ are combined to one factor which is calculated using the absolute intensity per decay as given in the ENSDF database. Isotopes which do not contribute by itself but feed another decay that contains a magnetic dipole transition are listed as well in this table.}\\
\label{Tab:reactorisotopes}

\endlastfoot

$^{89}$Rb - $\beta$-decay & 1473 & 0.91 & 0.91 & 3.84$\times$10$^{-3}$& 0.75 \\
 & 466 & 0.91 & 0.91 & 7.55$\times$10$^{-4}$& 0.18 \\
 & 1940 & 0.91 & 0.91 & 3.59$\times$10$^{-3}$& 0.45 \\
 & 2007 & 0.91 & 0.91 & 2.58$\times$10$^{-2}$ & -0.8 \\
 & 1419 & 0.91 & 0.91 & 1.01$\times$10$^{-3}$& -0.6 \\
 & 2451 & 0.91 & 0.91 & 5.66$\times$10$^{-4}$& 0.18 \\
$^{90}$Rb - $\beta$-decay & 1060 & 0.61 & 0.61 & 9.54$\times$10$^{-2}$ & 0.5 \\
$^{90}$Rb$^m$ - $\beta$-decay & 1060 & 1.60 & 1.60 & 7.61$\times$10$^{-2}$ & 0.5 \\
$^{91}$Rb - $\beta$-decay & 948 & 2.90 & 2.90 & 1.17$\times$10$^{-2}$ & n.a. \\
 & 1042 & 2.90 & 2.90 & 2.20$\times$10$^{-2}$ & n.a. \\
 & 1137 & 2.90 & 2.90 & 3.89$\times$10$^{-2}$ & n.a. \\
 & 93 & 2.90 & 2.90 & 3.38$\times$10$^{-1}$ & 3.3 \\
$^{92}$Rb - $\beta$-decay & 386 & 5.70 & 5.70 & 2.43$\times$10$^{-4}$& n.a. \\
 & 393 & 5.70 & 5.70 & 1.22$\times$10$^{-3}$& n.a. \\
 & 570 & 5.70 & 5.70 & 5.44$\times$10$^{-3}$& 0.21 \\
 & 756 & 5.70 & 5.70 & 1.09$\times$10$^{-3}$& -0.09 \\
 & 963 & 5.70 & 5.70 & 1.47$\times$10$^{-3}$& 1.7 \\
 & 1239 & 5.70 & 5.70 & 3.52$\times$10$^{-4}$& -3.3 \\
 & 1325 & 5.70 & 5.70 & 1.34$\times$10$^{-3}$& -0.27 \\
$^{92}$Rb - $\beta$n-decay & 0 & 5.70 & 0 & 0 & n.a. \\
$^{93}$Rb - $\beta$n-decay & 393 & 3.30 & 0.04 & 1.93$\times$10$^{-4}$& n.a. \\
 & 570 & 3.30 & 0.04 & 4.28$\times$10$^{-4}$& 0.21 \\
 & 963 & 3.30 & 0.04 & 2.35$\times$10$^{-4}$& 1.7 \\
$^{93}$Rb - $\beta$-decay & 213 & 3.30 & 3.26 & 7.76$\times$10$^{-2}$ & 2.4 \\
 & 219 & 3.30 & 3.26 & 3.19$\times$10$^{-2}$ & n.a. \\
 & 432 & 3.30 & 3.26 & 2.02$\times$10$^{-1}$ & n.a. \\
$^{93}$Sr - $\beta$-decay & 590 & 2.10 & 5.42 & 6.80$\times$10$^{-1}$ & n.a. \\
$^{94}$Rb - $\beta$n-decay & 219 & 1.17 & 0.07 & 7.14$\times$10$^{-3}$& n.a. \\
 & 432 & 1.17 & 0.07 & 4.20$\times$10$^{-2}$ & n.a. \\
$^{94}$Rb - $\beta$-decay & 503 & 1.17 & 1.10 & 1.22$\times$10$^{-2}$ & -0.35 \\
 & 677 & 1.17 & 1.10 & 2.56$\times$10$^{-2}$ & -0.54 \\
 & 1089 & 1.17 & 1.10 & 1.20$\times$10$^{-1}$ & 0.02 \\
 & 1577 & 1.17 & 1.10 & 2.23$\times$10$^{-1}$ & -0.02 \\
$^{94}$Sr - $\beta$-decay & 0 & 4.60 & 5.76 & 0 & n.a. \\
$^{95}$Rb - $\beta$n-decay & 0 & 0.69 & 0.06 & 0 & n.a. \\
$^{95}$Rb - $\beta$-decay & 256 & 0.69 & 0.63 & 9.80$\times$10$^{-4}$& n.a. \\
 & 328 & 0.69 & 0.63 & 9.31$\times$10$^{-2}$ & n.a. \\
 & 331 & 0.69 & 0.63 & 1.57$\times$10$^{-2}$ & n.a. \\
 & 352 & 0.69 & 0.63 & 4.90$\times$10$^{-1}$ & n.a. \\
 & 660 & 0.69 & 0.63 & 4.12$\times$10$^{-2}$ & n.a. \\
 & 680 & 0.69 & 0.63 & 1.48$\times$10$^{-1}$ & n.a. \\
 & 682 & 0.69 & 0.63 & 1.57$\times$10$^{-2}$ & n.a. \\
$^{95}$Sr - $\beta$-decay & 0 & 5.30 & 5.93 & 0 & n.a. \\
$^{96}$Sr - $\beta$-decay & 122 & 2.50 & 2.50 & 1.00 & n.a. \\
 & 652 & 2.50 & 2.50 & 1.00 & -0.11 \\
 & 213 & 2.50 & 2.50 & 1.06$\times$10$^{-2}$ & n.a. \\
 & 279.4 & 2.50 & 2.50 & 1.15$\times$10$^{-1}$ & -0.05 \\
$^{96}$Y$^m$ - $\beta$-decay & 174 & 0.94 & 0.94 & 2.11$\times$10$^{-2}$ & n.a. \\
 & 289 & 0.94 & 0.94 & 8.80$\times$10$^{-3}$& -0.4 \\
 & 475 & 0.94 & 0.94 & 3.08$\times$10$^{-2}$ & -0.09 \\
$^{96}$Y - $\beta$-decay & 475 & 0 & 2.50 & 1.88$\times$10$^{-3}$& -0.09 \\
 & 918 & 0 & 2.50 & 7.52$\times$10$^{-4}$& n.a. \\
$^{97}$Sr - $\beta$n-decay & 0 & 0.81 & 0 & 0 & n.a. \\
$^{97}$Sr - $\beta$-decay & 307 & 0.81 & 0.81 & 1.00$\times$10$^{-1}$ & n.a. \\
$^{97}$Y - $\beta$-decay & 0 & 1.40 & 2.21 & 0 & n.a. \\
$^{97}$Y$^m$ - $\beta$-decay & 0 & 2.00 & 2.00 & 0 & n.a. \\
$^{97}$Zr - $\beta$-decay & 254 & 0 & 4.24 & 1.15$\times$10$^{-2}$ & -0.04 \\
 & 355 & 0 & 4.24 & 2.09$\times$10$^{-2}$ & -0.04 \\
 & 602 & 0 & 4.24 & 1.38$\times$10$^{-2}$ & 0.11 \\
 & 703 & 0 & 4.24 & 1.01$\times$10$^{-2}$ & 0.19 \\
 & 1148 & 0 & 4.24 & 2.61$\times$10$^{-2}$ & 0.5 \\
$^{97}$Nb - $\beta$-decay & 657 & 0 & 4.24 & 9.82$\times$10$^{-1}$ & -0.05 \\
$^{98}$Y - $\beta$n-decay & 0 & 1.40 & 0 & 0 & n.a. \\
$^{98}$Y - $\beta$-decay & 0 & 1.40 & 1.40 & 0 & n.a. \\
$^{98}$Y$^m$ - $\beta$n-decay & 0 & 0.80 & 0.03 & 0 & n.a. \\
$^{98}$Y$^m$ - $\beta$-decay & 521 & 0.80 & 0.77 & 1.26$\times$10$^{-2}$ & 0.2 \\
$^{98}$Zr - $\beta$-decay & 0 & 0 & 2.18 & 0 & n.a. \\
$^{98}$Nb - $\beta$-decay & 0 & 0 & 2.18 & 0 & n.a. \\
$^{99}$Y - $\beta$n-decay & 0 & 1.20 & 0.01 & 0 & n.a. \\
$^{99}$Y - $\beta$-decay & 46 & 1.20 & 1.19 & 4.23$\times$10$^{-4}$& n.a. \\
 & 53.3 & 1.20 & 1.19 & 1.13$\times$10$^{-2}$ & n.a. \\
 & 66.6 & 1.20 & 1.19 & 2.59$\times$10$^{-3}$& n.a. \\
 & 82.2 & 1.20 & 1.19 & 3.62$\times$10$^{-3}$& n.a. \\
 & 90.4 & 1.20 & 1.19 & 5.50$\times$10$^{-3}$& n.a. \\
 & 121.7 & 1.20 & 1.19 & 4.70$\times$10$^{-1}$ & n.a. \\
 & 127.6 & 1.20 & 1.19 & 2.07$\times$10$^{-3}$& n.a. \\
 & 149 & 1.20 & 1.19 & 1.60$\times$10$^{-3}$& n.a. \\
 & 186 & 1.20 & 1.19 & 3.90$\times$10$^{-3}$& n.a. \\
 & 189 & 1.20 & 1.19 & 2.54$\times$10$^{-3}$& n.a. \\
 & 192 & 1.20 & 1.19 & 1.93$\times$10$^{-2}$ & n.a. \\
$^{99}$Zr - $\beta$-decay & 28.4 & 3.30 & 4.49 & 2.15$\times$10$^{-3}$& n.a. \\
 & 55.9 & 3.30 & 4.49 & 2.15$\times$10$^{-2}$ & n.a. \\
 & 81.8 & 3.30 & 4.49 & 3.03$\times$10$^{-2}$ & n.a. \\
 & 86.7 & 3.30 & 4.49 & 3.85$\times$10$^{-4}$& n.a. \\
 & 178.9 & 3.30 & 4.49 & 5.39$\times$10$^{-2}$ & n.a. \\
 & 347.5 & 3.30 & 4.49 & 4.40$\times$10$^{-4}$& n.a. \\
 & 387.4 & 3.30 & 4.49 & 7.98$\times$10$^{-2}$ & n.a. \\
 & 429.3 & 3.30 & 4.49 & 1.09$\times$10$^{-1}$ & n.a. \\
 & 461.8 & 3.30 & 4.49 & 5.50$\times$10$^{-1}$ & n.a. \\
$^{99}$Nb$^m$ - $\beta$-decay & 138 & 5.00 & 9.49 & 8.10$\times$10$^{-1}$ & n.a. \\
 & 197 & 5.00 & 9.49 & 9.72$\times$10$^{-4}$& n.a. \\
 & 208 & 5.00 & 9.49 & 4.05$\times$10$^{-4}$& n.a. \\
 & 251 & 5.00 & 9.49 & 1.30$\times$10$^{-3}$& n.a. \\
 & 253 & 5.00 & 9.49 & 3.32$\times$10$^{-3}$& n.a. \\
 & 264 & 5.00 & 9.49 & 3.89$\times$10$^{-3}$& n.a. \\
 & 277 & 5.00 & 9.49 & 4.05$\times$10$^{-4}$& n.a. \\
 & 408 & 5.00 & 9.49 & 4.05$\times$10$^{-4}$& n.a. \\
 & 451 & 5.00 & 9.49 & 7.61$\times$10$^{-3}$& n.a. \\
 & 462 & 5.00 & 9.49 & 2.43$\times$10$^{-3}$& n.a. \\
 & 518 & 5.00 & 9.49 & 8.91$\times$10$^{-4}$& n.a. \\
 & 527.9 & 5.00 & 9.49 & 6.48$\times$10$^{-4}$& n.a. \\
 & 548 & 5.00 & 9.49 & 2.19$\times$10$^{-3}$& n.a. \\
 & 600 & 5.00 & 9.49 & 1.15$\times$10$^{-2}$ & n.a. \\
 & 631 & 5.00 & 9.49 & 4.05$\times$10$^{-4}$& n.a. \\
 & 671 & 5.00 & 9.49 & 4.05$\times$10$^{-4}$& n.a. \\
 & 727 & 5.00 & 9.49 & 9.72$\times$10$^{-4}$& n.a. \\
 & 767 & 5.00 & 9.49 & 4.05$\times$10$^{-4}$& n.a. \\
 & 768 & 5.00 & 9.49 & 1.30$\times$10$^{-2}$ & n.a. \\
 & 907 & 5.00 & 9.49 & 1.11$\times$10$^{-2}$ & n.a. \\
 & 1044 & 5.00 & 9.49 & 3.16$\times$10$^{-3}$& n.a. \\
 & 1107 & 5.00 & 9.49 & 1.22$\times$10$^{-2}$ & n.a. \\
 & 1228 & 5.00 & 9.49 & 5.67$\times$10$^{-4}$& n.a. \\
$^{99}$Mo - $\beta$-decay & 41 & 0 & 9.49 & 1.04$\times$10$^{-2}$ & 0.008 \\
 & 140 & 0 & 9.49 & 3.05$\times$10$^{-5}$ & 0.129 \\
 & 366 & 0 & 9.49 & 1.20$\times$10$^{-2}$ & n.a. \\
 & 380 & 0 & 9.49 & 1.05$\times$10$^{-4}$& 1.3 \\
 & 410 & 0 & 9.49 & 1.95$\times$10$^{-5}$ & 0.5 \\
 & 457 & 0 & 9.49 & 8.17$\times$10$^{-5}$ & n.a. \\
 & 529 & 0 & 9.49 & 5.32$\times$10$^{-4}$& n.a. \\
$^{100}$Y - $\beta$-decay & 665 & 0.24 & 0.24 & 1.04$\times$10$^{-1}$ & n.a. \\
$^{100}$Zr - $\beta$-decay & 33 & 4.10 & 4.34 & 2.60$\times$10$^{-3}$& n.a. \\
 & 104 & 4.10 & 4.34 & 9.10$\times$10$^{-3}$& n.a. \\
$^{101}$Zr - $\beta$-decay & 108 & 2.80 & 1.00 & 1.22$\times$10$^{-3}$& n.a. \\
 & 119 & 2.80 & 1.00 & 1.10$\times$10$^{-1}$ & 0.33 \\
 & 136 & 2.80 & 1.00 & 1.04$\times$10$^{-2}$ & 1.05 \\
 & 140 & 2.80 & 1.00 & 1.89$\times$10$^{-2}$ & 0.25 \\
 & 186 & 2.80 & 1.00 & 3.05$\times$10$^{-3}$& n.a. \\
 & 205 & 2.80 & 1.00 & 6.10$\times$10$^{-2}$ & n.a. \\
$^{101}$Nb - $\beta$-decay & 13.5 & 0 & 2.80 & 2.10$\times$10$^{-1}$ & 0.03 \\
 & 43.5 & 0 & 2.80 & 2.10$\times$10$^{-1}$ & 1.06 \\
 & 114 & 0 & 2.80 & 4.20$\times$10$^{-3}$& n.a. \\
 & 118 & 0 & 2.80 & 2.94$\times$10$^{-2}$ & n.a. \\
 & 157 & 0 & 2.80 & 6.72$\times$10$^{-2}$ & n.a. \\
$^{101}$Mo - $\beta$-decay & 6 & 0 & 2.80 & 5.43$\times$10$^{-3}$& 0.01 \\
 & 9 & 0 & 2.80 & 2.11$\times$10$^{-2}$ & 0.015 \\
 & 80 & 0 & 2.80 & 3.73$\times$10$^{-2}$ & n.a. \\
$^{102}$Zr - $\beta$-decay & 64 & 3.90 & 3.90 & 8.60$\times$10$^{-2}$ & n.a. \\
 & 73 & 3.90 & 3.90 & 1.17$\times$10$^{-2}$ & n.a. \\
 & 85 & 3.90 & 3.90 & 7.00$\times$10$^{-3}$& n.a. \\
 & 96 & 3.90 & 3.90 & 1.10$\times$10$^{-2}$ & n.a. \\
 & 102 & 3.90 & 3.90 & 1.37$\times$10$^{-2}$ & n.a. \\
 & 136 & 3.90 & 3.90 & 1.40$\times$10$^{-2}$ & n.a. \\
 & 156 & 3.90 & 3.90 & 3.40$\times$10$^{-2}$ & n.a. \\
$^{102}$Nb - $\beta$-decay & 0 & 0 & 3.90 & 0 & n.a. \\
$^{102}$Mo - $\beta$-decay & 43 & 0 & 3.90 & 2.47$\times$10$^{-4}$& n.a. \\
 & 93 & 0 & 3.90 & 1.03$\times$10$^{-3}$& n.a. \\
 & 13 & 0 & 3.90 & 2.32$\times$10$^{-3}$& n.a. \\
 & 148 & 0 & 3.90 & 3.76$\times$10$^{-2}$ & n.a. \\
 & 211 & 0 & 3.90 & 3.80$\times$10$^{-2}$ & n.a. \\
 & 223 & 0 & 3.90 & 1.44$\times$10$^{-2}$ & n.a. \\
$^{102}$Tc - $\beta$-decay & 0 & 0 & 3.90 & 0 & n.a. \\
$^{136}$I - $\beta$-decay & 344 & 1.20 & 1.20 & 2.40$\times$10$^{-2}$ & n.a. \\
 & 1321 & 1.20 & 1.20 & 2.48$\times$10$^{-1}$ & n.a. \\
$^{136}$I$^m$ - $\beta$-decay & 369 & 1.07 & 1.07 & 1.75$\times$10$^{-1}$ & n.a. \\
 & 482 & 1.07 & 1.07 & 1.75$\times$10$^{-2}$ & n.a. \\
$^{140}$Cs - $\beta$-decay & 672 & 2.60 & 2.60 & 1.15$\times$10$^{-2}$ & 0.13 \\
 & 820 & 2.60 & 2.60 & 2.49$\times$10$^{-3}$& n.a. \\
 & 908 & 2.60 & 2.60 & 8.62$\times$10$^{-2}$ & -0.6 \\
 & 1008 & 2.60 & 2.60 & 8.25$\times$10$^{-3}$& -4.5 \\
 & 1129 & 2.60 & 2.60 & 1.22$\times$10$^{-2}$ & 1.7 \\
 & 1391 & 2.60 & 2.60 & 1.86$\times$10$^{-2}$ & 0.18 \\
 & 1634 & 2.60 & 2.60 & 2.53$\times$10$^{-2}$ & n.a. \\
 & 2101 & 2.60 & 2.60 & 3.03$\times$10$^{-2}$ & -0.09 \\
 & 2268 & 2.60 & 2.60 & 1.20$\times$10$^{-2}$ & -0.19 \\
$^{140}$Ba - $\beta$-decay & 13.8 & 0 & 2.60 & 1.22$\times$10$^{-2}$ & 0.01 \\
 & 30 & 0 & 2.60 & 1.41$\times$10$^{-1}$ & 0.009 \\
 & 63 & 0 & 2.60 & 2.93$\times$10$^{-7}$ & n.a. \\
 & 113 & 0 & 2.60 & 1.46$\times$10$^{-4}$& n.a. \\
 & 119 & 0 & 2.60 & 6.10$\times$10$^{-4}$& n.a. \\
 & 132 & 0 & 2.60 & 2.02$\times$10$^{-3}$& n.a. \\
 & 162 & 0 & 2.60 & 6.22$\times$10$^{-2}$ & 0.08 \\
 & 304 & 0 & 2.60 & 4.29$\times$10$^{-2}$ & 0.1 \\
 & 423 & 0 & 2.60 & 3.10$\times$10$^{-2}$ & n.a. \\
 & 437 & 0 & 2.60 & 1.93$\times$10$^{-2}$ & n.a. \\
 & 537 & 0 & 2.60 & 2.44$\times$10$^{-1}$ & n.a. \\
$^{140}$La - $\beta$-decay & 64 & 0 & 2.60 & 1.43$\times$10$^{-4}$& n.a. \\
 & 69 & 0 & 2.60 & 7.54$\times$10$^{-4}$& n.a. \\
 & 109 & 0 & 2.60 & 2.19$\times$10$^{-3}$& 0.26 \\
 & 131 & 0 & 2.60 & 4.67$\times$10$^{-3}$& -0.13 \\
 & 175 & 0 & 2.60 & 1.27$\times$10$^{-3}$& n.a. \\
 & 242 & 0 & 2.60 & 4.10$\times$10$^{-3}$& -0.6 \\
 & 266 & 0 & 2.60 & 4.67$\times$10$^{-3}$& -0.14 \\
 & 328 & 0 & 2.60 & 2.03$\times$10$^{-1}$ & -0.049 \\
 & 432 & 0 & 2.60 & 2.90$\times$10$^{-2}$ & -0.04 \\
 & 751 & 0 & 2.60 & 4.33$\times$10$^{-2}$ & 0.38 \\
 & 816 & 0 & 2.60 & 2.33$\times$10$^{-1}$ & -0.03 \\
 & 867 & 0 & 2.60 & 5.50$\times$10$^{-2}$ & n.a. \\
 & 925 & 0 & 2.60 & 6.90$\times$10$^{-2}$ & -0.22 \\
 & 951 & 0 & 2.60 & 5.19$\times$10$^{-3}$& 0.01 \\
 & 2.547 & 0 & 2.60 & 1.01$\times$10$^{-3}$& n.a. \\
$^{141}$Cs - $\beta$n-decay & 0 & 4.30 & 0 & 0 & n.a. \\
$^{141}$Cs - $\beta$-decay & 6.5 & 4.30 & 4.30 & 1.62$\times$10$^{-3}$& n.a. \\
 & 48.5 & 4.30 & 4.30 & 7.90$\times$10$^{-2}$ & 0.36 \\
$^{141}$Ba - $\beta$-decay & 113 & 1.70 & 6.00 & 1.00$\times$10$^{-2}$ & n.a. \\
 & 114 & 1.70 & 6.00 & 1.05$\times$10$^{-3}$& n.a. \\
 & 163 & 1.70 & 6.00 & 2.91$\times$10$^{-3}$& n.a. \\
 & 181 & 1.70 & 6.00 & 5.10$\times$10$^{-3}$& n.a. \\
 & 1000 & 1.70 & 6.00 & 8.65$\times$10$^{-2}$ & n.a. \\
 & 277 & 1.70 & 6.00 & 2.32$\times$10$^{-1}$ & n.a. \\
 & 304 & 1.70 & 6.00 & 2.52$\times$10$^{-1}$ & 1.8 \\
 & 458 & 1.70 & 6.00 & 4.96$\times$10$^{-2}$ & n.a. \\
$^{141}$La - $\beta$-decay & 0 & 0 & 6.00 & 0 & n.a. \\
$^{141}$Ce - $\beta$-decay & 145 & 0 & 6.00 & 4.84$\times$10$^{-1}$ & 0.069 \\
$^{142}$Cs - $\beta$n-decay & 0 & 2.20 & 0 & 0 & n.a. \\
$^{142}$Cs - $\beta$-decay & 1064 & 2.20 & 2.20 & 8.91$\times$10$^{-3}$& 10 \\
$^{142}$Ba - $\beta$-decay & 64 & 3.70 & 5.92 & 9.06$\times$10$^{-4}$& n.a. \\
 & 68 & 3.70 & 5.92 & 7.83$\times$10$^{-4}$& n.a. \\
 & 70 & 3.70 & 5.92 & 2.64$\times$10$^{-3}$& n.a. \\
 & 77 & 3.70 & 5.92 & 9.52$\times$10$^{-2}$ & n.a. \\
 & 123 & 3.70 & 5.92 & 9.27$\times$10$^{-3}$& n.a. \\
 & 231 & 3.70 & 5.92 & 1.22$\times$10$^{-1}$ & -0.16 \\
 & 255 & 3.70 & 5.92 & 2.06$\times$10$^{-1}$ & -0.26 \\
 & 269 & 3.70 & 5.92 & 9.27$\times$10$^{-3}$& n.a. \\
 & 286 & 3.70 & 5.92 & 1.11$\times$10$^{-2}$ & n.a. \\
 & 309 & 3.70 & 5.92 & 2.60$\times$10$^{-2}$ & -0.74 \\
 & 364 & 3.70 & 5.92 & 4.74$\times$10$^{-2}$ & -0.77 \\
 & 425 & 3.70 & 5.92 & 5.75$\times$10$^{-2}$ & 0.31 \\
 & 457 & 3.70 & 5.92 & 3.75$\times$10$^{-3}$& n.a. \\
 & 604 & 3.70 & 5.92 & 4.20$\times$10$^{-3}$& n.a. \\
$^{142}$La - $\beta$-decay & 861.6 & 0 & 5.92 & 7.21$\times$10$^{-4}$& 0.03 \\
 & 895 & 0 & 5.92 & 3.63$\times$10$^{-3}$& -0.63 \\
 & 962 & 0 & 5.92 & 1.65$\times$10$^{-4}$& -0.56 \\
 & 1130 & 0 & 5.92 & 2.06$\times$10$^{-4}$& -6 \\
 & 1363 & 0 & 5.92 & 9.27$\times$10$^{-4}$& 0.16 \\
$^{143}$Cs - $\beta$n-decay & 0 & 1.30 & 0.02 & 0 & n.a. \\
$^{143}$Cs - $\beta$-decay & 146 & 1.30 & 1.28 & 4.03$\times$10$^{-3}$& n.a. \\
 & 160 & 1.30 & 1.28 & 4.41$\times$10$^{-3}$& n.a. \\
 & 195 & 1.30 & 1.28 & 1.26$\times$10$^{-1}$ & n.a. \\
 & 229 & 1.30 & 1.28 & 2.52$\times$10$^{-2}$ & 0.6 \\
 & 232 & 1.30 & 1.28 & 8.32$\times$10$^{-2}$ & n.a. \\
 & 263 & 1.30 & 1.28 & 3.65$\times$10$^{-2}$ & n.a. \\
 & 273 & 1.30 & 1.28 & 4.28$\times$10$^{-2}$ & n.a. \\
 & 306 & 1.30 & 1.28 & 6.80$\times$10$^{-2}$ & n.a. \\
$^{143}$Ba - $\beta$-decay & 177 & 3.40 & 4.68 & 1.21$\times$10$^{-2}$ & n.a. \\
 & 178 & 3.40 & 4.68 & 3.03$\times$10$^{-2}$ & n.a. \\
 & 182 & 3.40 & 4.68 & 7.65$\times$10$^{-3}$& n.a. \\
 & 208 & 3.40 & 4.68 & 1.08$\times$10$^{-2}$ & n.a. \\
 & 211 & 3.40 & 4.68 & 2.50$\times$10$^{-1}$ & 0.07 \\
 & 254 & 3.40 & 4.68 & 2.43$\times$10$^{-2}$ & n.a. \\
 & 261 & 3.40 & 4.68 & 1.68$\times$10$^{-2}$ & n.a. \\
 & 291 & 3.40 & 4.68 & 8.13$\times$10$^{-2}$ & 0.99 \\
 & 398 & 3.40 & 4.68 & 1.46$\times$10$^{-1}$ & n.a. \\
 & 431 & 3.40 & 4.68 & 2.78$\times$10$^{-2}$ & n.a. \\
$^{143}$La - $\beta$-decay & 0 & 2.90 & 7.58 & 0 & n.a. \\
$^{143}$Ce - $\beta$-decay & 57 & 0 & 7.58 & 1.17$\times$10$^{-1}$ & 0.039 \\
 & 231 & 0 & 7.58 & 2.05$\times$10$^{-2}$ & n.a. \\
 & 293 & 0 & 7.58 & 4.28$\times$10$^{-1}$ & 0.77 \\
 & 371 & 0 & 7.58 & 2.48$\times$10$^{-4}$& n.a. \\
 & 389 & 0 & 7.58 & 3.64$\times$10$^{-4}$& n.a. \\
 & 433 & 0 & 7.58 & 1.59$\times$10$^{-3}$& n.a. \\
 & 446 & 0 & 7.58 & 1.50$\times$10$^{-4}$& n.a. \\
 & 490 & 0 & 7.58 & 2.16$\times$10$^{-2}$ & n.a. \\
 & 559 & 0 & 7.58 & 3.17$\times$10$^{-4}$& n.a. \\
 & 587 & 0 & 7.58 & 2.67$\times$10$^{-3}$& n.a. \\
 & 664 & 0 & 7.58 & 5.69$\times$10$^{-2}$ & n.a. \\
 & 722 & 0 & 7.58 & 5.39$\times$10$^{-2}$ & n.a. \\
 & 806 & 0 & 7.58 & 2.87$\times$10$^{-4}$& n.a. \\
 & 880 & 0 & 7.58 & 1.03$\times$10$^{-2}$ & n.a. \\
 & 1002 & 0 & 7.58 & 7.53$\times$10$^{-4}$& n.a. \\
 & 1031 & 0 & 7.58 & 2.01$\times$10$^{-4}$& n.a. \\
 & 1046 & 0 & 7.58 & 1.20$\times$10$^{-4}$& n.a. \\
 & 1060 & 0 & 7.58 & 3.64$\times$10$^{-4}$& n.a. \\
 & 1103 & 0 & 7.58 & 4.15$\times$10$^{-3}$& n.a. \\
$^{143}$Pr - $\beta$-decay & 0 & 0 & 7.58 & 0 & n.a. \\
$^{144}$Ba - $\beta$-decay & 16 & 4.00 & 4.00 & 5.34$\times$10$^{-3}$& n.a. \\
 & 42 & 4.00 & 4.00 & 1.37$\times$10$^{-2}$ & n.a. \\
 & 69 & 4.00 & 4.00 & 2.75$\times$10$^{-2}$ & n.a. \\
 & 82 & 4.00 & 4.00 & 5.98$\times$10$^{-2}$ & n.a. \\
 & 104 & 4.00 & 4.00 & 2.31$\times$10$^{-1}$ & n.a. \\
 & 111 & 4.00 & 4.00 & 5.50$\times$10$^{-2}$ & n.a. \\
 & 115 & 4.00 & 4.00 & 2.68$\times$10$^{-2}$ & 1.09 \\
 & 173 & 4.00 & 4.00 & 1.52$\times$10$^{-1}$ & n.a. \\
 & 182 & 4.00 & 4.00 & 1.04$\times$10$^{-2}$ & n.a. \\
 & 228 & 4.00 & 4.00 & 1.31$\times$10$^{-2}$ & 1.1 \\
 & 295 & 4.00 & 4.00 & 9.24$\times$10$^{-3}$& n.a. \\
$^{144}$La - $\beta$-decay & 453 & 0.40 & 4.40 & 1.89$\times$10$^{-2}$ & -0.45 \\
 & 890 & 0.40 & 4.40 & 1.26$\times$10$^{-2}$ & 0.68 \\
 & 969 & 0.40 & 4.40 & 3.31$\times$10$^{-2}$ & 0.5 \\
 & 978 & 0.40 & 4.40 & 1.90$\times$10$^{-2}$ & -0.32 \\
 & 1082 & 0.40 & 4.40 & 5.47$\times$10$^{-3}$& -5.6 \\
 & 1092 & 0.40 & 4.40 & 1.01$\times$10$^{-2}$ & 5 \\
 & 1102 & 0.40 & 4.40 & 1.25$\times$10$^{-2}$ & -0.63 \\
 & 1347 & 0.40 & 4.40 & 1.50$\times$10$^{-2}$ & -0.09 \\
 & 1358 & 0.40 & 4.40 & 2.83$\times$10$^{-3}$& n.a. \\
 & 1422 & 0.40 & 4.40 & 1.13$\times$10$^{-2}$ & -3.5 \\
 & 1524 & 0.40 & 4.40 & 3.48$\times$10$^{-2}$ & n.a. \\
 & 1624 & 0.40 & 4.40 & 6.98$\times$10$^{-3}$& 0.13 \\
 & 1632 & 0.40 & 4.40 & 1.04$\times$10$^{-2}$ & 0.53 \\
 & 1714 & 0.40 & 4.40 & 9.81$\times$10$^{-3}$& n.a. \\
 & 1755 & 0.40 & 4.40 & 1.06$\times$10$^{-2}$ & n.a. \\
 & 1942 & 0.40 & 4.40 & 1.40$\times$10$^{-2}$ & 0.07 \\
 & 2352 & 0.40 & 4.40 & 4.72$\times$10$^{-3}$& n.a. \\
$^{144}$Ce - $\beta$-decay & 33 & 0 & 4.40 & 2.00$\times$10$^{-3}$& n.a. \\
 & 41 & 0 & 4.40 & 2.57$\times$10$^{-3}$& 0.042 \\
 & 53 & 0 & 4.40 & 9.98$\times$10$^{-4}$& n.a. \\
 & 80 & 0 & 4.40 & 1.36$\times$10$^{-2}$ & n.a. \\
 & 133 & 0 & 4.40 & 1.11$\times$10$^{-1}$ & n.a. \\
$^{144}$Pr - $\beta$-decay & 864 & 0 & 4.40 & 2.42$\times$10$^{-5}$ & -0.75 \\
$^{145}$Ba - $\beta$-decay & 66 & 1.80 & 1.80 & 5.27$\times$10$^{-2}$ & n.a. \\
 & 92 & 1.80 & 1.80 & 7.31$\times$10$^{-2}$ & n.a. \\
 & 97 & 1.80 & 1.80 & 1.70$\times$10$^{-1}$ & n.a. \\
 & 123 & 1.80 & 1.80 & 1.19$\times$10$^{-2}$ & n.a. \\
 & 162 & 1.80 & 1.80 & 3.74$\times$10$^{-2}$ & n.a. \\
 & 186 & 1.80 & 1.80 & 1.87$\times$10$^{-2}$ & n.a. \\
$^{145}$La - $\beta$-decay & 48 & 3.80 & 5.60 & 1.44$\times$10$^{-2}$ & n.a. \\
 & 70 & 3.80 & 5.60 & 1.08$\times$10$^{-1}$ & n.a. \\
 & 118 & 3.80 & 5.60 & 3.61$\times$10$^{-2}$ & n.a. \\
 & 164 & 3.80 & 5.60 & 2.70$\times$10$^{-2}$ & n.a. \\
 & 170 & 3.80 & 5.60 & 3.19$\times$10$^{-2}$ & n.a. \\
$^{145}$Ce - $\beta$-decay & 62 & 0 & 5.60 & 1.33$\times$10$^{-1}$ & n.a. \\
 & 126 & 0 & 5.60 & 4.90$\times$10$^{-3}$& n.a. \\
 & 208 & 0 & 5.60 & 1.09$\times$10$^{-2}$ & n.a. \\
 & 284 & 0 & 5.60 & 8.14$\times$10$^{-2}$ & n.a. \\
 & 351 & 0 & 5.60 & 2.54$\times$10$^{-2}$ & n.a. \\
 & 423 & 0 & 5.60 & 3.84$\times$10$^{-2}$ & n.a. \\
$^{145}$Pr - $\beta$-decay & 72 & 0 & 5.60 & 2.61$\times$10$^{-3}$& n.a. \\
$^{146}$La - $\beta$-decay & 0 & 1.30 & 1.30 & 0 & n.a. \\
$^{146}$Ce - $\beta$-decay & 12 & 0 & 1.30 & 1.71$\times$10$^{-3}$& n.a. \\
 & 23 & 0 & 1.30 & 7.74$\times$10$^{-3}$& n.a. \\
 & 35 & 0 & 1.30 & 1.11$\times$10$^{-2}$ & 0.87 \\
 & 52 & 0 & 1.30 & 1.02$\times$10$^{-2}$ & n.a. \\
 & 87 & 0 & 1.30 & 6.86$\times$10$^{-3}$& n.a. \\
 & 98 & 0 & 1.30 & 3.82$\times$10$^{-2}$ & n.a. \\
 & 106 & 0 & 1.30 & 6.14$\times$10$^{-3}$& n.a. \\
 & 134 & 0 & 1.30 & 8.23$\times$10$^{-2}$ & n.a. \\
 & 141 & 0 & 1.30 & 3.40$\times$10$^{-2}$ & 0.6 \\
 & 251 & 0 & 1.30 & 2.72$\times$10$^{-2}$ & n.a. \\
$^{146}$Pr - $\beta$-decay & 191 & 0 & 1.30 & 1.48$\times$10$^{-4}$& n.a. \\
 & 508 & 0 & 1.30 & 4.44$\times$10$^{-3}$& n.a. \\
 & 736 & 0 & 1.30 & 7.22$\times$10$^{-2}$ & -0.07 \\
 & 849 & 0 & 1.30 & 7.59$\times$10$^{-4}$& n.a. \\
 & 1017 & 0 & 1.30 & 1.19$\times$10$^{-2}$ & -13 \\
 & 1323 & 0 & 1.30 & 5.37$\times$10$^{-3}$& 4.6 \\
 & 1333 & 0 & 1.30 & 6.67$\times$10$^{-3}$& 1.4 \\
 & 1452 & 0 & 1.30 & 2.20$\times$10$^{-2}$ & 0.68 \\
 & 1524 & 0 & 1.30 & 1.50$\times$10$^{-1}$ & 0.03 \\
 & 1690 & 0 & 1.30 & 5.97$\times$10$^{-3}$& n.a. \\
 & 2149 & 0 & 1.30 & 2.32$\times$10$^{-4}$& n.a. \\
$^{147}$La - $\beta$n-decay & 0 & 1.60 & 0 & 0 & n.a. \\
$^{147}$La - $\beta$-decay & 59 & 1.60 & 1.60 & 5.28$\times$10$^{-3}$& n.a. \\
 & 69 & 1.60 & 1.60 & 6.69$\times$10$^{-3}$& n.a. \\
 & 118 & 1.60 & 1.60 & 1.76$\times$10$^{-1}$ & n.a. \\
 & 215 & 1.60 & 1.60 & 4.05$\times$10$^{-2}$ & n.a. \\
 & 235 & 1.60 & 1.60 & 3.63$\times$10$^{-2}$ & n.a. \\
$^{147}$Ce - $\beta$-decay & 93 & 0 & 1.60 & 6.42$\times$10$^{-2}$ & n.a. \\
 & 178 & 0 & 1.60 & 3.96$\times$10$^{-3}$& n.a. \\
 & 198 & 0 & 1.60 & 2.96$\times$10$^{-2}$ & n.a. \\
 & 219 & 0 & 1.60 & 2.24$\times$10$^{-2}$ & 0.57 \\
 & 247 & 0 & 1.60 & 7.57$\times$10$^{-3}$& 0.4 \\
 & 254 & 0 & 1.60 & 1.02$\times$10$^{-2}$ & n.a. \\
 & 289 & 0 & 1.60 & 1.88$\times$10$^{-2}$ & n.a. \\
 & 292 & 0 & 1.60 & 6.07$\times$10$^{-3}$& n.a. \\
 & 802 & 0 & 1.60 & 1.01$\times$10$^{-2}$ & n.a. \\
$^{147}$Pr - $\beta$-decay & 49.9 & 0 & 1.60 & 4.19$\times$10$^{-2}$ & 0.42 \\
 & 78 & 0 & 1.60 & 9.91$\times$10$^{-2}$ & 0.48 \\
 & 100 & 0 & 1.60 & 5.21$\times$10$^{-3}$& n.a. \\
 & 128 & 0 & 1.60 & 7.23$\times$10$^{-2}$ & 0.4 \\
 & 141 & 0 & 1.60 & 1.32$\times$10$^{-2}$ & n.a. \\
 & 148 & 0 & 1.60 & 6.43$\times$10$^{-4}$& n.a. \\
 & 168 & 0 & 1.60 & 4.34$\times$10$^{-4}$& n.a. \\
 & 202 & 0 & 1.60 & 2.96$\times$10$^{-3}$& n.a. \\
 & 249 & 0 & 1.60 & 1.37$\times$10$^{-2}$ & 0.9 \\
 & 335 & 0 & 1.60 & 4.54$\times$10$^{-2}$ & 3.6 \\
 & 389 & 0 & 1.60 & 1.41$\times$10$^{-2}$ & 0.82 \\
 & 463 & 0 & 1.60 & 1.45$\times$10$^{-4}$& n.a. \\
 & 467 & 0 & 1.60 & 1.87$\times$10$^{-2}$ & n.a. \\
 & 503 & 0 & 1.60 & 3.62$\times$10$^{-3}$& n.a. \\
 & 516 & 0 & 1.60 & 1.69$\times$10$^{-2}$ & n.a. \\
 & 581 & 0 & 1.60 & 2.68$\times$10$^{-3}$& n.a. \\
 & 631 & 0 & 1.60 & 6.15$\times$10$^{-3}$& n.a. \\
 & 1112 & 0 & 1.60 & 1.23$\times$10$^{-3}$& n.a. \\
$^{147}$Nd - $\beta$-decay & 81 & 0 & 1.60 & 9.09$\times$10$^{-6}$ & n.a. \\
 & 91 & 0 & 1.60 & 2.81$\times$10$^{-1}$ & 0.094 \\
 & 120 & 0 & 1.60 & 3.76$\times$10$^{-3}$& 0.05 \\
 & 149 & 0 & 1.60 & 3.88$\times$10$^{-5}$ & n.a. \\
 & 155 & 0 & 1.60 & 4.14$\times$10$^{-5}$ & n.a. \\
 & 191 & 0 & 1.60 & 3.74$\times$10$^{-5}$ & n.a. \\
 & 196 & 0 & 1.60 & 1.90$\times$10$^{-3}$& -0.2 \\
 & 272 & 0 & 1.60 & 1.32$\times$10$^{-4}$& 0.1 \\
 & 275 & 0 & 1.60 & 9.10$\times$10$^{-3}$& 0.107 \\
 & 319 & 0 & 1.60 & 2.13$\times$10$^{-2}$ & -0.37 \\
 & 398 & 0 & 1.60 & 9.09$\times$10$^{-3}$& 0.3 \\
 & 408 & 0 & 1.60 & 1.87$\times$10$^{-4}$& 0.57 \\
 & 440 & 0 & 1.60 & 1.28$\times$10$^{-2}$ & 0.62 \\
 & 489 & 0 & 1.60 & 1.55$\times$10$^{-3}$& 1.2 \\
 & 531 & 0 & 1.60 & 1.34$\times$10$^{-1}$ & -0.4 \\
 & 589 & 0 & 1.60 & 3.88$\times$10$^{-4}$& n.a. \\
 & 594 & 0 & 1.60 & 2.83$\times$10$^{-3}$& 0.55 \\
 & 680 & 0 & 1.60 & 2.94$\times$10$^{-4}$& n.a. \\
 & 686 & 0 & 1.60 & 8.86$\times$10$^{-3}$& -0.95 s
\end{longtable}
\end{center}

\end{document}